\documentclass[useAMS,referee]{ajhg} 
\usepackage{graphicx}
\usepackage{amsmath}
\usepackage{bbm}
\usepackage{amsfonts}
\usepackage{textcomp}
\usepackage{subfigure}
\usepackage[super, sort]{cite}
\usepackage{color}
\usepackage{booktabs}
\usepackage{threeparttable}

\newcommand{\bs}{\boldsymbol}
\newcommand{\iid}{\stackrel{\mathrm{iid}}{\sim}}
\newcommand{\Prob}{\textsf{Prob}}

\def\urltilda{\kern -.15em\lower .7ex\hbox{\~{}}\kern .04em}
\def\urldot{\kern -.10em.\kern -.10em}
\def\urlhttp{http\kern -.10em\lower -.1ex\hbox{:}\kern -.12em\lower 0ex\hbox{/}\kern -.18em\lower 0ex\hbox{/}}

\title[Generalized Admixture Mapping for Complex Traits]{Generalized Admixture Mapping for Complex Traits}
\author
{Bin Zhu$^{1,2,*}$, Allison E. Ashley-Koch$^{1}$ and David B. Dunson$^{2}$ \\
$^1$ Center for Human Genetics, Duke University Medical Center, Durham, North Carolina 27710, U.S.A. \\
$^2$ Department of Statistical Science, Duke University, Durham,  North Carolina 27708, U.S.A.\\
$^*$ Correspondence: bin.zhu@duke.edu
}

\begin{document}

\date{}

\pagerange{}  
\volume{}
\pubyear{}
\artmonth{}

\doi{}

\begin{titlepage}

\end{titlepage}
\maketitle
\newpage

\section*{Abstract}
Admixture mapping is a popular tool to identify regions of the genome associated with traits in a recently admixed population. Existing methods have been developed primarily for identification of a single locus influencing a dichotomous trait within a case-control study design. We propose a generalized admixture mapping (GLEAM) approach, a flexible and powerful regression method for both quantitative and qualitative traits, which is able to test for association between the trait and local ancestries in multiple loci simultaneously and adjust for covariates. The new method is based on the generalized linear model and utilizes a quadratic normal moment prior to incorporate admixture prior information. Through simulation, we demonstrate that GLEAM achieves lower type I error rate and higher power than existing methods both for qualitative traits and more significantly for quantitative traits. We applied GLEAM to genome-wide SNP data from the Illumina African American panel derived from a cohort of black woman participating in the Healthy Pregnancy, Healthy Baby study and identified a locus on chromosome $2$ associated with the averaged maternal mean arterial pressure during 24 to 28 weeks of pregnancy. 
\newpage


\section*{Introduction}
Admixture mapping, also known as mapping by admixture linkage disequilibrium (MALD), has become an important tool for localizing disease genes.
A number of admixture mapping studies, focused on primarily on African American populations, have successfully identified candidate loci associated with common complex traits and biomarkers. Examples include hypertension \cite{zhu2005admixture,zhu2007admixture}, multiple sclerosis \cite{reich2005whole}, cardiovascular disease \cite{reiner2005population}, prostate cancer \cite{freedman2006admixture,bock2009results}, interleukin 6 levels \cite{reich2007admixture},  end-stage renal disease \cite{kao2008myh9,kopp2008myh9}, white blood cell counts \cite{nalls2008admixture}, blood lipid levels \cite{basu2009admixture}, obesity \cite{cheng2009admixture}, retinal vascular caliber \cite{cheng2010admixture},  peripheral arterial disease \cite{scherer2010admixture}, blood pressure \cite{Zhu01062011} and acute lymphoblastic leukemia \cite{yang2011ancestry}. Among these new found susceptibility loci, the association between end-stage renal disease and the region harboring MYH9 gene has been reported by multiple independent studies \cite{kao2008myh9,kopp2008myh9,ashleymyh9}. The 8q24 prostate cancer locus\cite{freedman2006admixture} has been confirmed by a series of follow-up admixture mapping and genome-wide association studies (GWAS) \cite{gudmundsson2007genome,haiman2007multiple,schumacher2007common,wang2007two,bock2009results}; and the locus on 5p13 contributing to inter-individual blood
pressure variation \cite{zhu2005admixture} has been verified by multiple large-scale GWAS. \cite{Zhu01062011}         

\begin{figure}
  \centering
\subfigure[]{\includegraphics[width=0.48\textwidth]{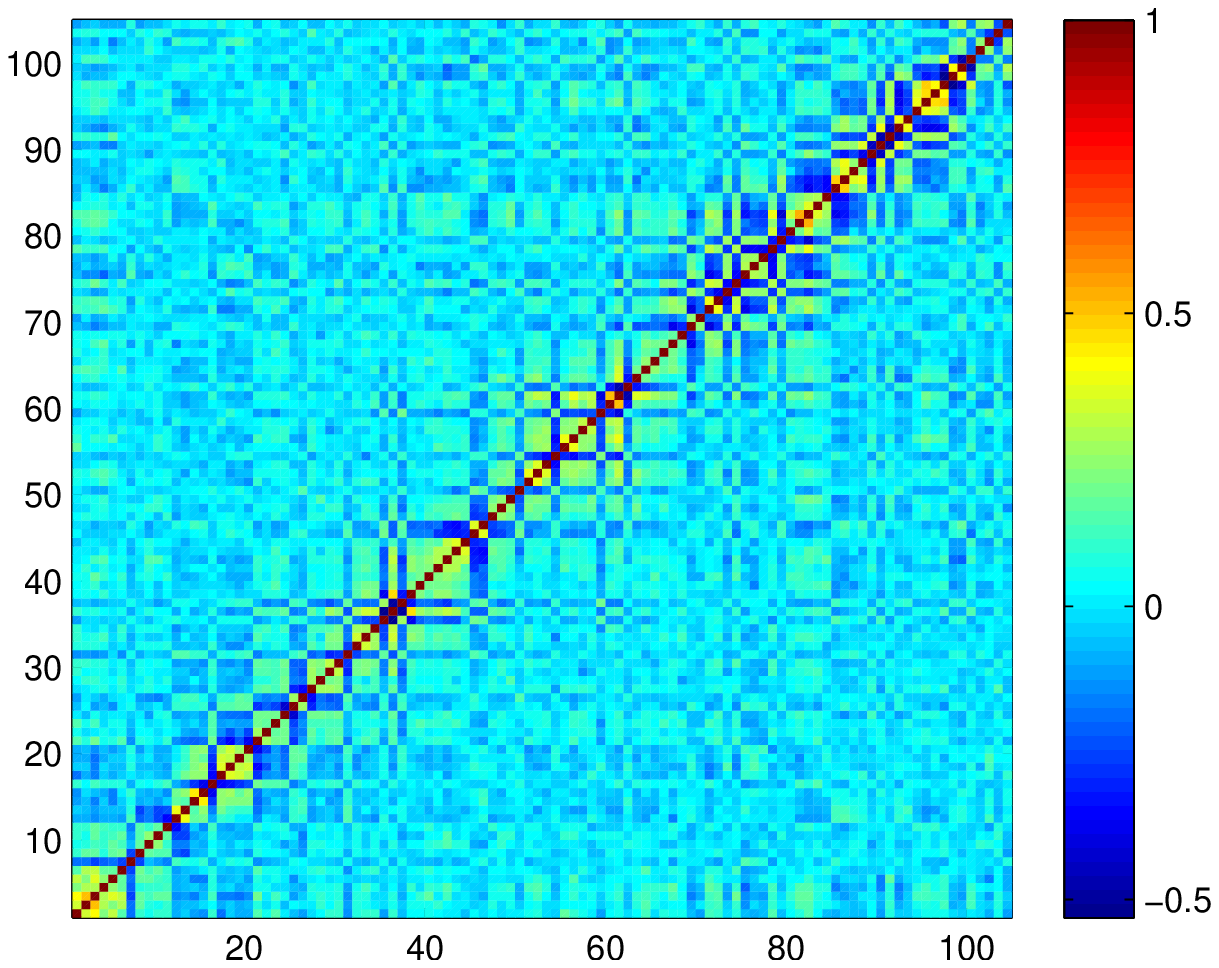}}
\subfigure[]{\includegraphics[width=0.48\textwidth]{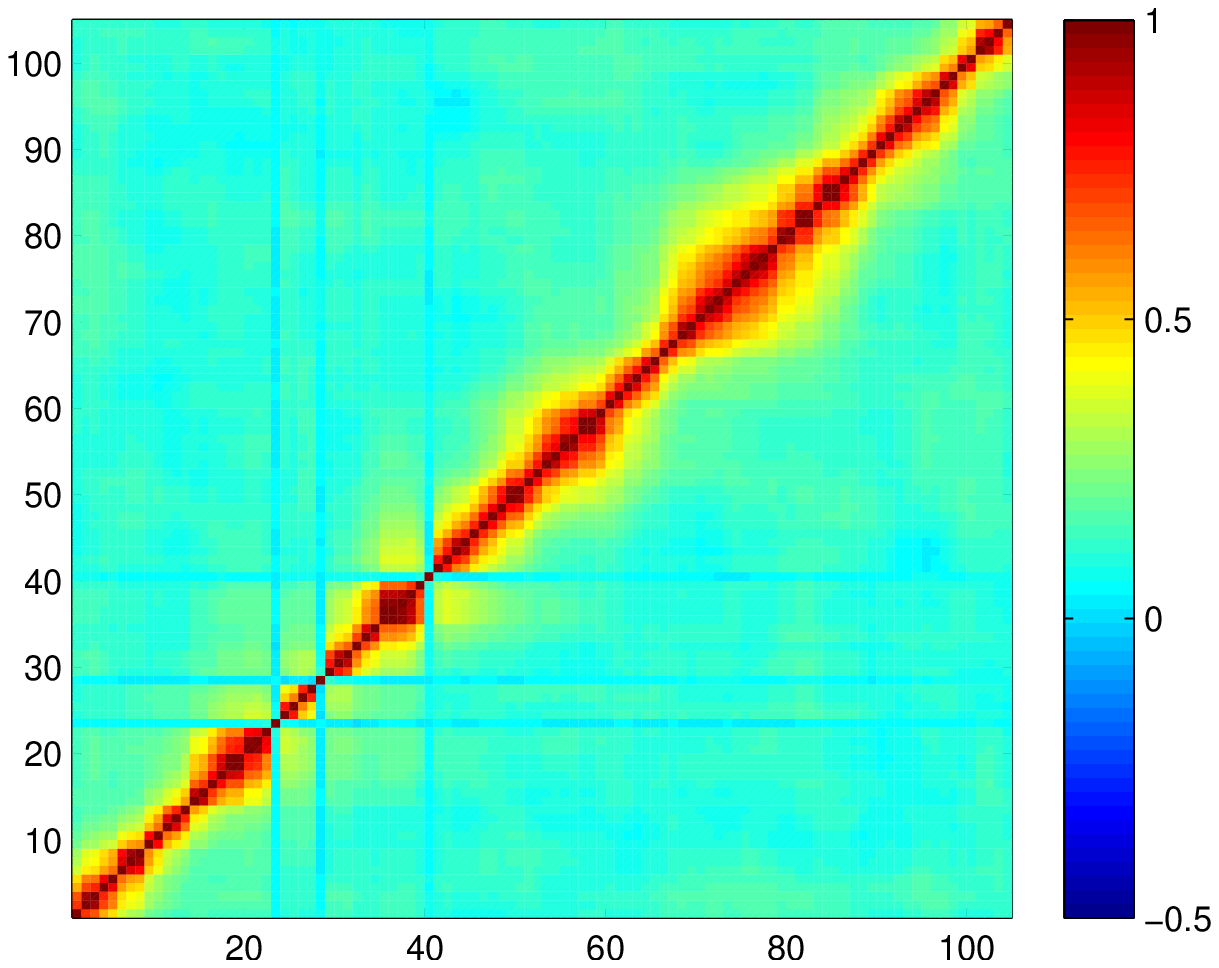}}\\
  \caption{ Heatmap of linkage disequilibrium in the chromosome 1 of 1001 African Americans. (a) Haplotype linkage disequilibrium, measured by correlation coefficients for the number of minor allele between pairs of loci; (b) Admixture linkage disequilibrium, measured by correlation coefficients for the local ancestry, i.e. number of Africa ancestry allele between pairs of loci, which are inferred using the Hidden Markov Model. \label{fig:ALD} }
\end{figure}

Admixture mapping is a genome-wide association approach to identify  susceptibility loci which confer risk or are linked with other loci harboring risk variants for complex-traits which have different prevalences between ancestral populations. \cite{stephens1994mapping,mckeigue2005prospects,reich2005will,
smith2005mapping,darvasi2005,zhu2008admixture,winkler2010admixture,chanock2011twist} In recently admixed populations, such as African Americans or Hispanic Americans, the chromosome resembles a mosaic of ancestry blocks, with alleles inherited together from one ancestral population within each block. The ancestral populations have different risks for the trait, which is assumed to be due in part to frequency differences in risk variants. For the ancestry block containing the risk variant, it is more likely to have originated from the high risk ancestral population than the low risk ancestral population. Hence, detecting the association between ancestry block and trait helps us to localize the  susceptibility loci. The ancestral status of a block at a specific genomic region, or local ancestry, is unobserved and can be estimated based on ancestry informative markers (AIMs), such as single nucleotide polymorphisms (SNPs), which vary in frequency across ancestral populations. AIMs tag the status of an ancestry block, similar to that of tagSNPs, which are used to characterize common haplotypes in a chromosomal region. In the African American population, the linkage disequilibrium due to admixture extends for a much wider region than the linkage disequilibrium between haplotypes \cite{smith2004high,patterson2004methods}, which is also illustrated in Figure \ref{fig:ALD}. Hence, compared to the tagSNP-based GWAS, admixture mapping requires many fewer markers to tag the whole genome and therefore increases the detection power at a reduced resolution, which is still higher than linkage analysis. \cite{patterson2004methods, smith2005mapping,winkler2010admixture} Moreover, admixture mapping is less vulnerable to allelic heterogeneity \cite{weiss2000many,darvasi2005}, since it relies on local ancestry instead of alleles directly.

Given the local ancestries of each individual \cite{falush2003inference, tang2006reconstructing,zhu2006classical, sankararaman2008estimating, price2009sensitive}, several hypothesis testing-based approaches have been proposed to test, one locus at a time, the null hypothesis that the AIM is unlinked to the complex-trait/disease. McKeigue \cite{mckeigue1997mapping} first applied the transmission-disequilibrium test \cite{ewens1995transmission} to explore the excess transmission of a risk variant from the high risk ancestral population at an AIM locus, and later \cite{mckeigue1998mapping} proposed a test for gametic disequilibrium between an AIM locus and the trait locus, conditional on the parental admixture. Patterson et al. \cite{patterson2004methods} suggested a Bayesian likelihood ratio test, comparing the likelihood under the alternative hypothesis (a given AIM locus is associated with the trait) versus the one under the null hypothesis, for cases and controls respectively. Zhu et al. \cite{zhu2004linkage} described a Z-score statistic, similar to one proposed by Montana and Pritchard \cite{montana2004statistical},  for testing the estimated local ancestry proportion is equal to one under the null hypothesis for case-control and case-only studies.  

Although considerable research has been devoted to single locus admixture mapping focused on dichotomous traits, less attention has been paid to admixture mapping for quantitative traits and to considering multiple loci simultaneously while adjusting for other risk factors. Quantitative traits have been the focus in many admixture mapping studies, such as lnterleukin 6 levels as inflammatory biomarkers for cardiovascular disease risk,\cite{reich2007admixture} ankle-arm index for peripheral arterial disease,\cite{scherer2010admixture} central retinal artery equivalent level for  retinal vascular caliber, \cite{cheng2010admixture} and white cell count for acute inflammation \cite{nalls2008admixture}. To apply existing admixture methods, the common practice has been to dichotomize subjects with the lowest and highest $q\%$ (e.g. $20\%$) of the quantitative trait value as cases and controls. The remaining subjects with in-between quantitative trait values are discarded,\cite{reich2007admixture,scherer2010admixture,cheng2010admixture} resulting in reduced power. In addition, complex traits are commonly caused by joint effects of the multiple genes and other risk factors, such as age, sex and smoking status. Investigating the association between AIM loci and a trait, one locus a time, without considering other loci or risk factors may capture a rather small proportion of joint effects and will possibly lead to inconsistent conclusions.\cite{zhu2005admixture,zhu2007admixture,deo2007high}   

With these motivations we propose regression-based generalized admixture mapping (GLEAM) for both quantitative and qualitative traits with the ability to examine the association between the complex trait and single or multiple loci simultaneously while also adjusting for other risk factors. The new approach is based on generalized linear models (GLMs) \cite{mccullagh1989generalized}, with linear regression for continuous traits, logistic regression for binary (e.g. case-control) traits and Poisson regression for count traits. The predictors in GLM include local ancestries at the given AIM loci and other risk factors. The local ancestry is defined as the number of alleles from the high risk ancestral population, for example, 0, 1 or 2 alleles from African ancestry at a given AIM locus. The association examined in  GLEAM can be adjusted by other risk factors. A related approach has been considered by Hoggart et al. \cite{hoggart2004design} for single locus without adjustment for other factors.  We assume for complex genetic traits that most loci have no association with the trait, a few loci may have small to modest association (e.g. odds ratio $<$ 2 for binary traits), and the loci with higher proportions of disease-causing alleles from the high-risk population would possibly have stronger association with the traits. \cite{smith2005mapping,winkler2010admixture} This prior knowledge is incorporated into GLEAM by using a quadratic normal moment (QNM) prior \cite{johnson2010use} for the coefficients in GLM (See more details in ``Material and Methods'' section) with the benefit of reducing the type I error while increasing the power, as demonstrated by the simulations in ``Results'' section.   

The number of AIMs ($1500\sim3000$)\cite{smith2004high} is usually larger than the number of study subjects, and keeps increasing ( $>$4000)\cite{tian2006genomewide,Tandon2011} with advances due to the HapMap project \cite{gibbs2003international,altshuler2005haplotype} and commercially available genome-wide SNP arrays.  It is not feasible to consider loci all together simultaneously due to the ``curse of dimensionality''. Rather, we propose a two-stage approach: in the first stage, we examine the association between local ancestries with the trait for one locus at a time and select a small subset of susceptibility loci; in the second stage, the associations between the various combinations of these selected loci and the trait are evaluated and the most significant ones are reported.  The associations in both steps are assessed by the Bayes factor (BF), the ratio between the likelihood of observed traits under the alternative hypothesis (presence of association between single or multiple loci with traits) and that under the null hypothesis (lack of association).\cite{kass1995bayes, stephens2009bayesian} 

The local ancestries are unobserved and will be inferred based on the AIMs using the Hidden Markov Model (HMM), \cite{rabiner1989tutorial, scott2002bayesian} with the focus on two-population admixture similar to that of Falush et al. \cite{falush2003inference} and Patterson et al. \cite{patterson2004methods} with one key difference: the recombination process is modeled non-parametrically. 
At each AIM locus, the number of alleles from the high risk ancestral population will be imputed multiple times for every subject, using an Markov chain Monte Carlo (MCMC) algorithm. Existing approaches only record the imputed frequency of the number of alleles from the high risk ancestral population individually \cite{falush2003inference} or across the population without accounting for imputation uncertainty \cite{patterson2004methods}. In contrast, our approach imputes multiple datasets of local ancestries, from which we are able to assess the association between the traits and local ancestries directly, while taking imputation uncertainty into account through Bayesian averaging. Importantly, the admixture linkage disequilibrium between the AIM loci is preserved in our multiple imputation approach, which is crucial for multilocus admixture mapping.

The remainder of paper is organized as follows. In Material and Methods, we first present the HMM for imputing the local ancestries, followed by the specification of the generalized linear model for quantitative and qualitative traits with QNM prior density. In Results, through the simulations we show that the new approach increases the power of admixture mapping while reducing the type I error rates compared to the popular method by Patterson et al. \cite{patterson2004methods} \hspace{-5pt}.  The new approach is applied to data from a large cohort study, the Healthy Pregnancy, Healthy Baby (HPHB) Study, and further extensions are considered in Discussion section.                       

\section*{Material and Methods}
\subsection*{Hidden Markov Model}
For a population-based design, suppose we have $I$ unrelated subjects, each of which has the same set of $J$ AIMs recorded. The local ancestry is measured by $S_{ij} \in \{0,1,2\}$, the  number of alleles from the high risk population A (e.g. African) for the $i$th subject and the $j$th AIM. $S_{ij}$ is unknown and will be imputed using the HMM. For African Americans with African and European ancestral populations, HMM assumes that given the $S_{ij}$, the distribution of $X_{ij} \in \{0,1,2\}$, the number of variant alleles, is independent of other $S_{ij^\prime}$ and $X_{ij^\prime}$ with $j^\prime \neq j$ and is specified by the observation probability mass matrix $\bs{P}_j=\{p_j(m,n)\}_{3\times 3}$ with  $p_j(m,n)=\Prob(X_{ij}=n \mid S_{ij}=m)$ and
{\small
\begin{eqnarray*}
\bs{P}_j=\bordermatrix{
    & X_{ij}=0 & X_{ij}=1 & X_{ij}=2\cr
  S_{ij}=0 \;\;& (1-p^B_j)(1-p^B_j) & 2p^B_j(1-p^B_j) & p^B_jp^B_j\cr
  S_{ij}=1 \;\;& (1-p^A_j)(1-p^B_j) & p^A_j(1-p^B_j)+p^B_j(1-p^A_j) & p^A_jp^B_j\cr
  S_{ij}=2 \;\;& (1-p^A_j)(1-p^A_j) & 2p^A_j(1-p^A_j) & p^A_jp^A_j\cr
   },
\end{eqnarray*}
}
where $p_j^A$ is the minor allele probability at loci $j$ in the high risk population $A$ and$p_j^B$ is the corresponding probability in the low risk population $B$.

The latent states $\bs{S}_i=\{S_{ij}\}_{1\times J}$, tagging the status of the ancestry blocks, are unobserved and modeled by an Markov chain which considers the genetic recombination events. Let $\rho_i$ denote the genome-wide proportion of alleles from the high risk population A for subject $i$,  $\bs{Q}_{i0}=[(1-\rho_i)^2,2\rho_i(1-\rho_i),\rho_i^2]^\prime$  initial state vector, $R_{ij} \in \{0,1,2\}$  the number of recombination events between AIM loci $j-1$ and $j$, $\bs{Q}^{(r)}_i=\{q^{(r)}_{i}(m,n)\}_{3 \times 3}$ the conditional state transition matrix given $r$ recombination events between the neighboring AIM loci with $q^{(r)}_{i}(m,n)=\Prob\left(S_{ij}=n \mid S_{i(j-1)}=m,  R_{ij}=r\right)$. The Markov chain $\bs{S}_i$ is governed by the state transition matrix $\bs{Q}_{ij}=\{q_{ij}(m,n)\}_{3 \times 3}$ with $q_{ij}(m,n)=\Prob\left(S_{ij}=n \mid S_{i(j-1)}=m\right)$.  $\bs{Q}_{ij} = \sum^2_{r=0} \bs{Q}^{(r)}_i\Prob(R_{ij}=r)$, where $\bs{Q}^{(0)}_i$, $\bs{Q}^{(1)}_i$ and  $\bs{Q}^{(2)}_i$ are specified as
{\small  
\begin{align*}
\bs{Q}^{(0)}_{i}&=\bordermatrix{%
&S_{ij}=0&S_{ij}=1&S_{ij}=2\cr
S_{i(j-1)}=0\;\;&1&0&0\cr
S_{i(j-1)}=1\;\;&0&1&0\cr
S_{i(j-1)}=2\;\;&0&0&1\cr
},\\
\bs{Q}^{(1)}_{i}&=\bordermatrix{%
&S_{ij}=0&S_{ij}=1&S_{ij}=2\cr
S_{i(j-1)}=0\;\;&1-\rho_i&\rho_i&0\cr
S_{i(j-1)}=1\;\;&\frac{1}{2}(1-\rho_i)&\frac{1}{2}&\frac{1}{2}\rho_i\cr
S_{i(j-1)}=2\;\;&0&1-\rho_i&\rho_i\cr
},\\
\bs{Q}^{(2)}_{i}&=\bordermatrix{%
&S_{ij}=0&S_{ij}=1&S_{ij}=2\cr
S_{i(j-1)}=0\;\;&(1-\rho_i)^2&2\rho_i(1-\rho_i)&\rho_i^2\cr
S_{i(j-1)}=1\;\;&(1-\rho_i)^2&2\rho_i(1-\rho_i)&\rho_i^2\cr
S_{i(j-1)}=2\;\;&(1-\rho_i)^2&2\rho_i(1-\rho_i)&\rho_i^2\cr
},
\end{align*} 
}
and $R_{ij} \sim \textsf{Bin}(2, \gamma_j)$ a binomial distribution with $\gamma_j$ the probability that a recombination event occurs between the neighboring AIM loci in a single chromosome.
Consequently, we can get,
\begingroup
\small
\begin{equation*}
\hspace{-20pt}\bs{Q}_{ij}=\bordermatrix{%
&S_{ij}=0&S_{ij}=1&S_{ij}=2\cr
S_{i(j-1)}=0\;\;&(1-\gamma_{j}\rho_i)^2&2\gamma_{j}\rho_i(1-\gamma_{j}\rho_i)&\gamma^2_{j}\rho^2_i\cr
S_{i(j-1)}=1\;\;&\gamma_{j}(1-\rho_i)(1-\gamma_{j}\rho_i)&\{1-\gamma_{j}(1-\rho_i)\}(1-\gamma_{j}\rho_i)+\gamma^2_{j}\rho_i(1-\rho_i)&\gamma_{j}\rho_i\{1-\gamma_{j}(1-\rho_i)\}\cr
S_{i(j-1)}=2\;\;&\gamma^2_{j}(1-\rho_i)^2&2\gamma_{j}(1-\rho_i)\{1-\gamma_{j}(1-\rho_i)\}&\{1-\gamma_{j}(1-\rho_i)\}^2\cr
}\vspace{10pt}
\end{equation*}
\endgroup 

We further specify informative prior distributions for the parameters $p^A_j$, $p^B_j$, $\gamma_j$ and $\rho_i$ involved in the HMM. Although the $p^A_j$ of the high risk population A is unknown, we have information on $p^A_{0j}$, the proportion of the variant allele $j$ in a subpopulation of high risk population A (e.g. YRI for African), from the HapMap or 1000 genome projects. Hence, we expect that $p^A_j$ would be close to $p^A_{0j}$ and specify $p^A_j \sim \textsf{Beta}\left(\tau^A p^A_{0j}, \tau^A(1-p^A_{0j}) \right)$ with the expectation $E(p^A_j)=p^A_{0j}$ and $\tau^A \sim \textsf{U}[50,1000]$ a uniform distribution to reflect the uncertainty in borrowing the subpopulation information. A similar specification is chosen for $p^B_j$ based on the proportion of the variant allele $j$ in a subpopulation of low risk population B (e.g. CEU for European). As for $\gamma_j$, it is well known that the recombination probability is roughly proportional to $d_j$ the genetic distance between $(j-1)$th and $j$th AIM loci. A common choice is $\gamma_j=1-\exp(-\lambda d_j)$ with $\lambda=6$ the number of recombination events per Morgan since admixture. \cite{patterson2004methods,falush2003inference} However, recombination `hotspots' can occur along the chromosomes where the recombination probabilities are much higher than the other regions \cite{stumpf2003estimating,myers2005fine,kong2010fine,Hinch2011}. For this reason, we avoid the above parametric specification of $\gamma_j$. Instead, we let $\gamma_j \sim \textsf{Beta}\left( \tau^\gamma \gamma_{0j}, \tau^\gamma (1-\gamma_{0j})  \right)$ with the expectation $E(\gamma_j)=\gamma_{0j}=1-\exp(-\lambda d_j)$. Hence, on average the probability of recombination is proportional to the genetic distance while allowing significant deviation (e.g. `hotspots' ) from the average. The deviation is measured by $\tau^\gamma$ with $Var(\gamma_j)=\frac{\gamma_{0j}(1-\gamma_{0j})}{\tau^\gamma+1}=\mu_0$. Additionally, for the admixed population, we often have knowledge about the proportions of ancestral populations at the population level. For example, the African American population in general consists of $80\%$ African ancestral population and $20\%$ European ancestral population. \cite{ smith2005mapping,winkler2010admixture} We borrow this population level information to specify $\rho_i$,  the subject specific proportion of high risk population A,  by letting $\rho_i \sim \textsf{Beta}\left(\tau^\rho \rho_{0i}, \tau^\rho (1-\rho_{0i})\right)$ with $\rho_{0i}$ (e.g. $0.8$ for African American) and $Var(\rho_i)=\frac{\rho_{0i}(1-\rho_{0i})}{\tau^\rho+1}=\nu_0$.  

We use an MCMC algorithm to sample the local ancestries $\bs{S}_i$ for $i=1,2,\dots,I$, along with other parameters. The details of MCMC are given in the Appendix.                    

\subsection*{Generalized linear model with QNM prior}
GLEAM is a regression method that extends the current approaches in various ways. The most obvious extension is to accommodate both quantitative and qualitative traits $\mathrm{y}_i$ through a generalized linear model with the ability to adjust for covariates $\bs{E}_i=(E_{i1},E_{i2},\dots, E_{iq})^\prime$. Specifically, we use the liner model for continuous traits, 
\begin{equation}
\label{eq:LinearModel}
\mathrm{y}_i=\beta_0+\bs{\beta}^\prime \bs{S}_i + \bs{\alpha}^\prime \bs{E}_i + \varepsilon_i,
\end{equation}
and the logistic model for dichotomous traits,
\begin{equation}
\label{eq:LogitModel}
\textsf{logit}\{\textsf{Prob}(\mathrm{y}_i=1)\}=\beta_0+\bs{\beta}^\prime \bs{S}_i + \bs{\alpha}^\prime \bs{E}_i,
\end{equation}
where $p$ local ancestries $\bs{S}_i=(S_{i1},S_{i2},\dots, S_{ip})^\prime$ are considered and centered to have mean zero, $\bs{\beta}=(\beta_{1},\beta_{2},\dots, \beta_{p})^\prime$ and $\bs{\alpha}=(\alpha_{1},\alpha_{2},\dots, \alpha_{q})^\prime$ are the regression coefficients for $\bs{S}_i$ and $\bs{E}_i$ respectively, and $\varepsilon_i \iid \textsf{N}(0,\sigma^2)$. We use the Bayes factor to assess the admixture association between local ancestries and the trait of interest. The Bayes factor is the ratio between the likelihood of observing the trait under the alternative hypothesis $\textsf{H}_1: \beta_1 \neq 0,\beta_2 \neq 0, \dots, \beta_p \neq 0$ and the likelihood under the null hypothesis $\textsf{H}_0: \beta_1=\beta_2=\dots=\beta_p=0$.

\begin{figure}
  \centering
\subfigure[]{\includegraphics[width=0.48\textwidth]{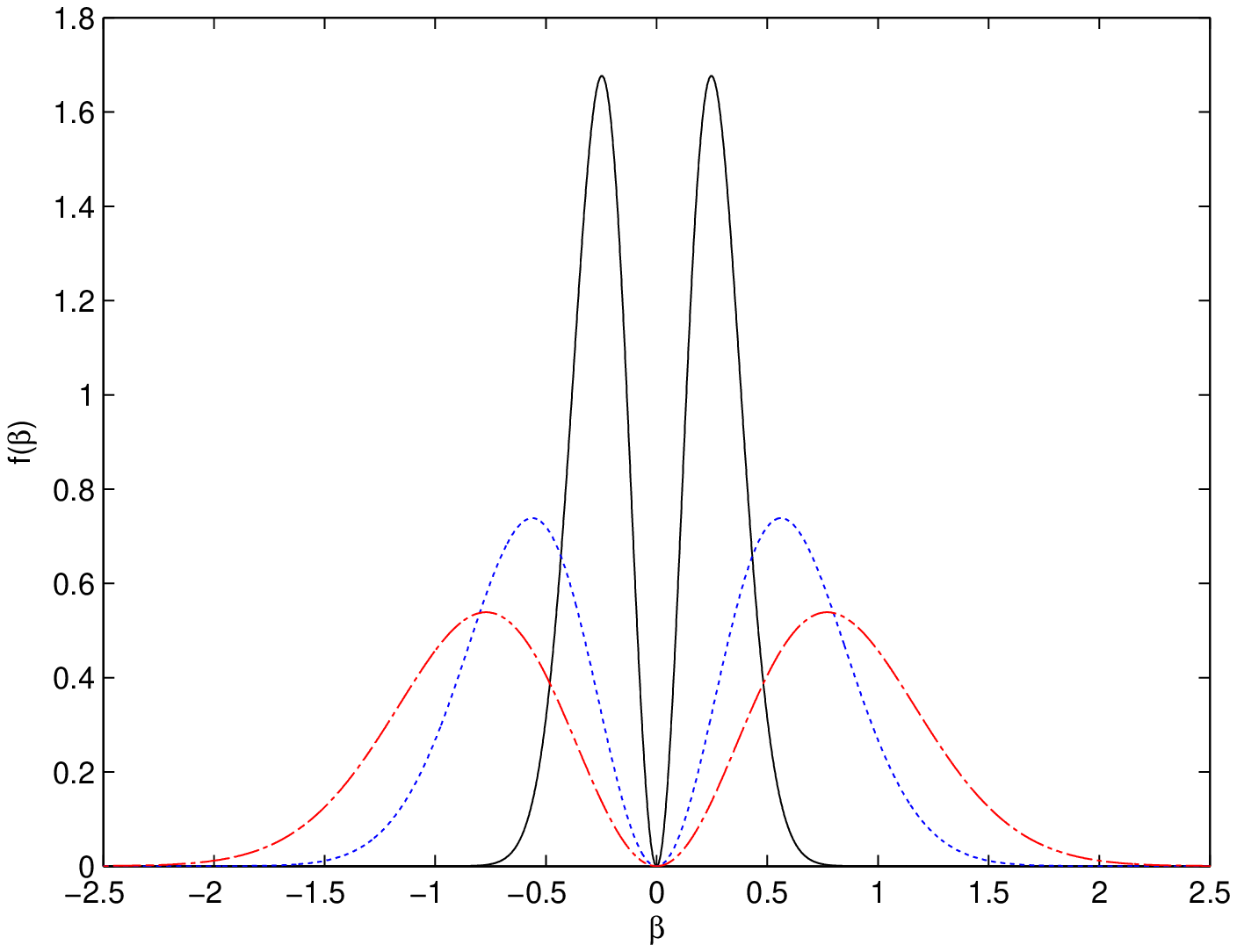}}
\subfigure[]{\includegraphics[width=0.48\textwidth]{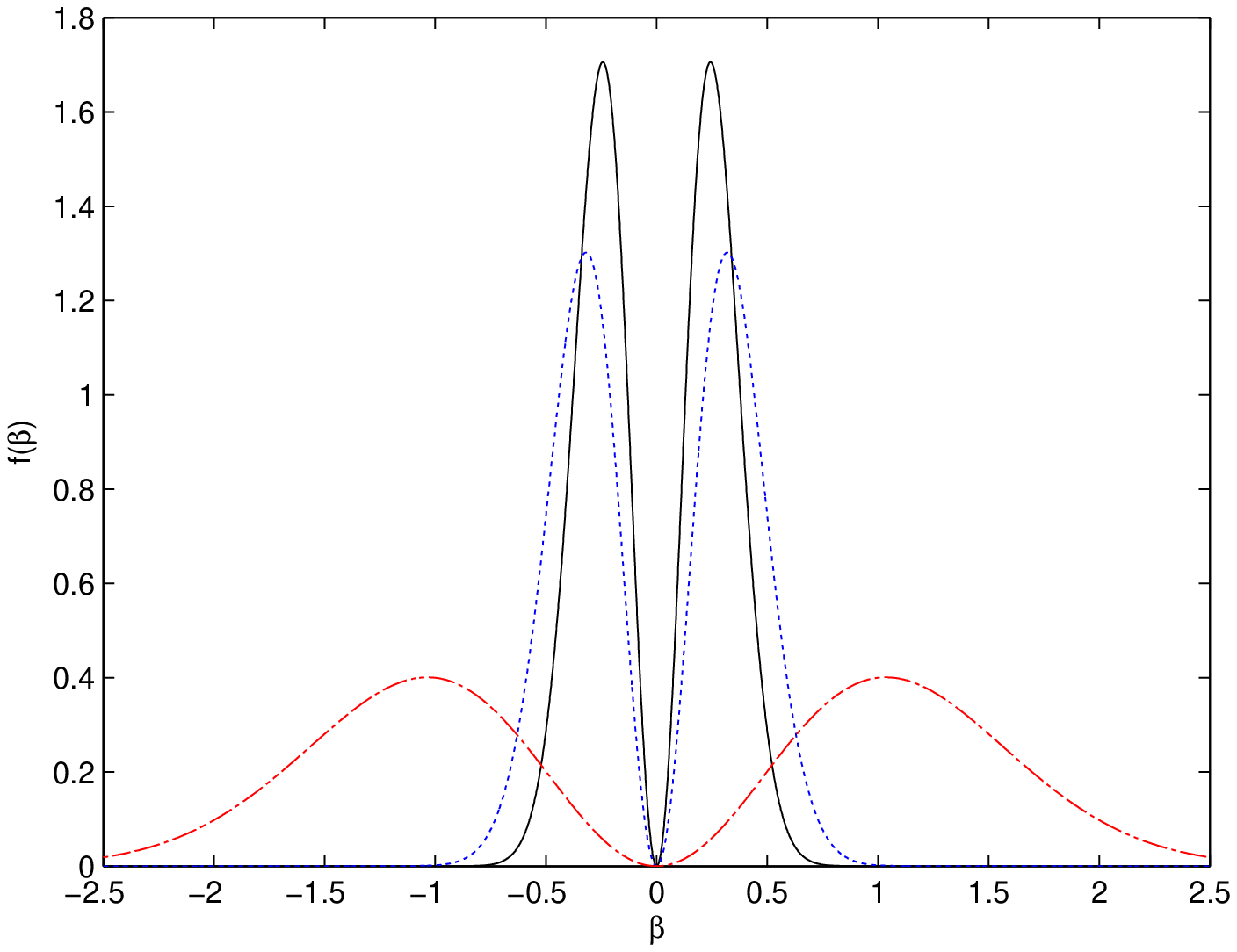}}
  \caption{ Univariate quadratic normal moment prior (a) for $\tau=0.01$ (\textemdash), $\tau=0.05$ ({\color{blue} $\cdotp \cdotp \cdotp$}) ,$\tau=0.1$ ({\color{red} $-\;\cdotp-$}) when $p_a=0.8$; (b) for $p_a=0.8$  (\textemdash), $p_a=0.9$ ({\color{blue} $\cdotp \cdotp \cdotp$}) and $p_a=0.99$ ({\color{red} $-\;\cdotp-$}), when $\tau=0.01$. In both cases, $\sigma^2=1$ and $\bs{\Sigma} = \left(\sum_{i=1}^{1000} S_i^2\right)^{-1}$ with $\textsf{Pr}(S_i=0)=(1-p_a)^2$, $\textsf{Pr}(S_i=1)=2p_a(1-p_a)$ and $\textsf{Pr}(S_i=2)=p_a^2$.
\label{fig:uni_QNMGP}}
\end{figure}

\begin{figure}
  \centering
\subfigure[]{\includegraphics[width=0.48\textwidth]{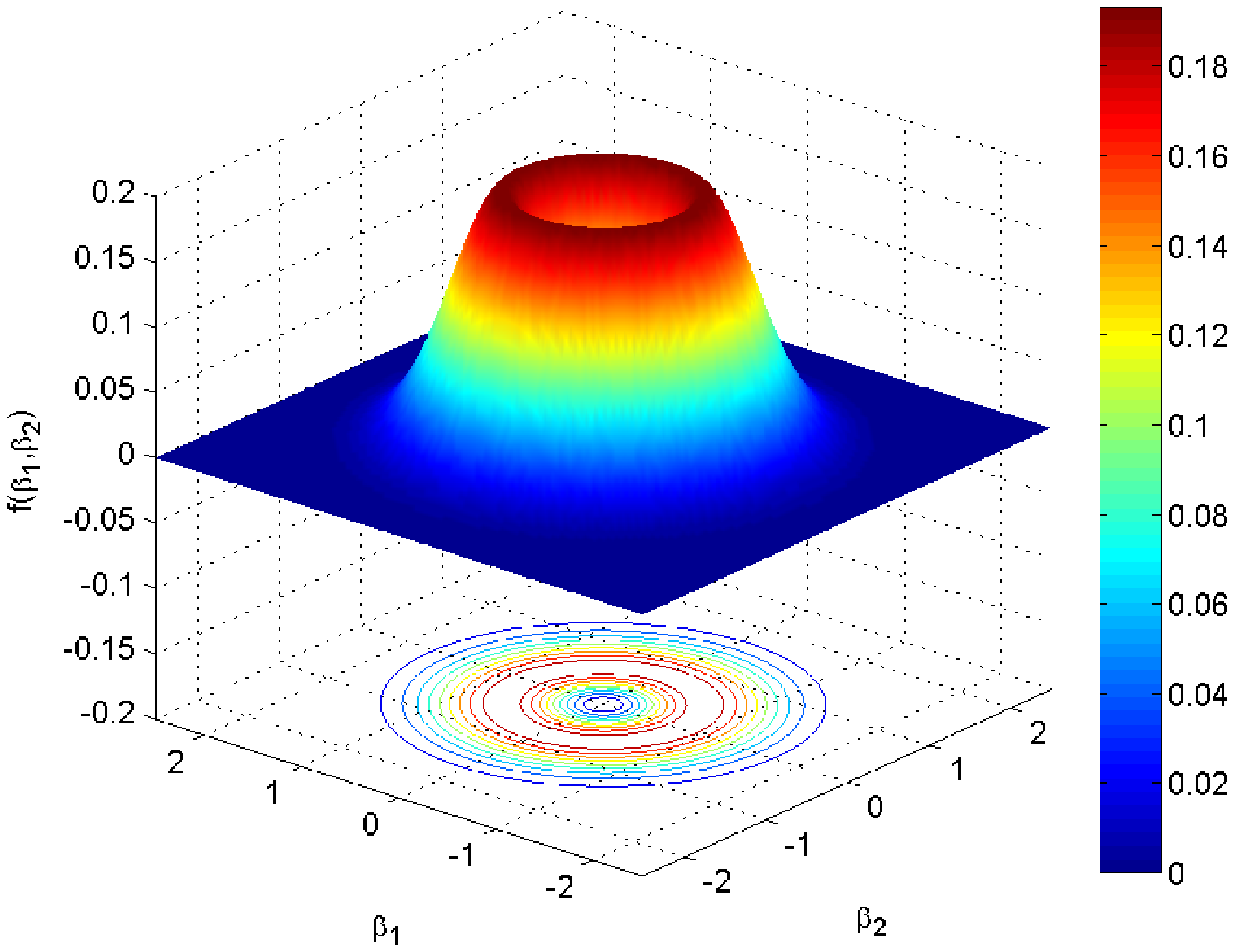}}
\subfigure[]{\includegraphics[width=0.48\textwidth]{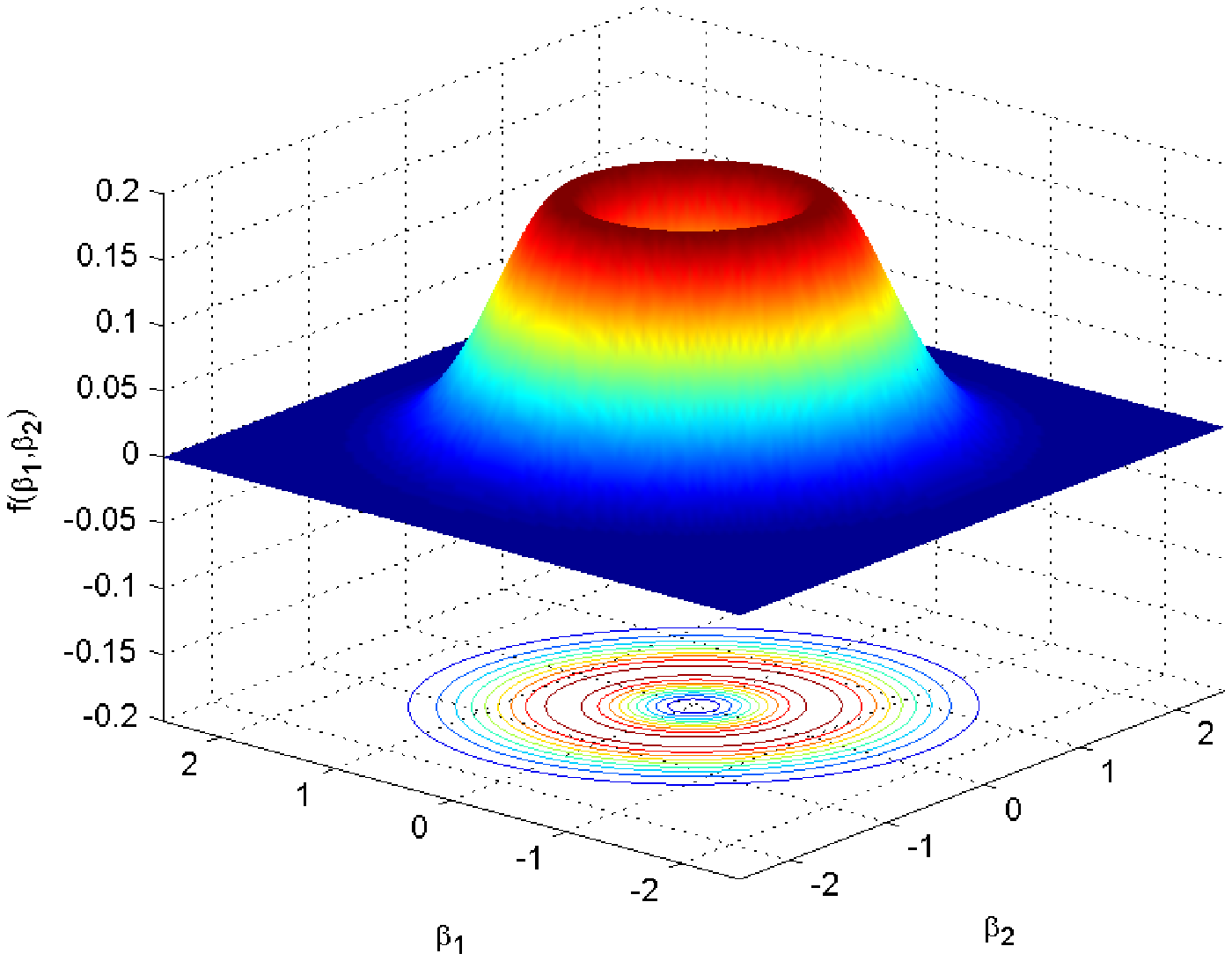}}\\
\subfigure[]{\includegraphics[width=0.48\textwidth]{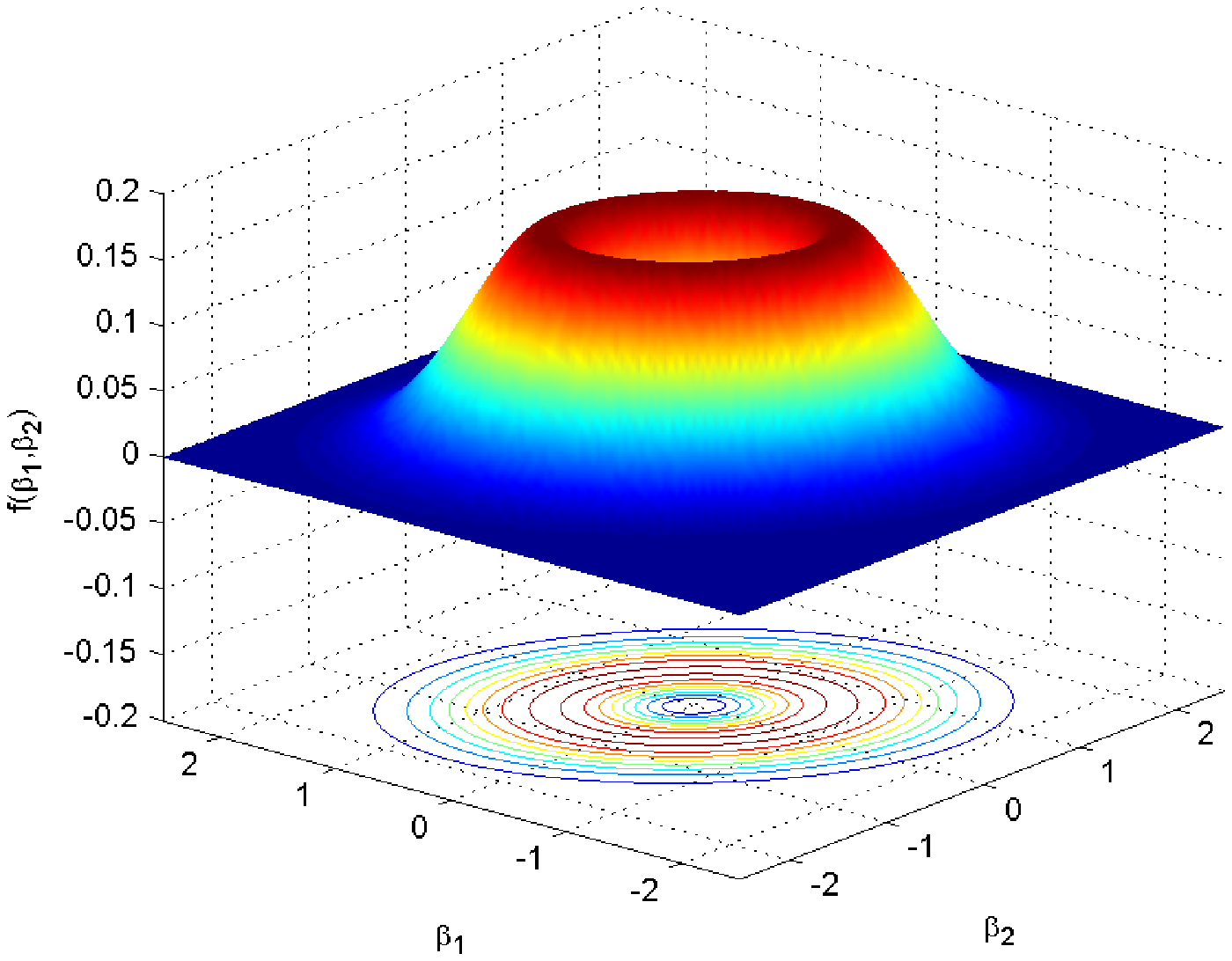}}
\subfigure[]{\includegraphics[width=0.48\textwidth]{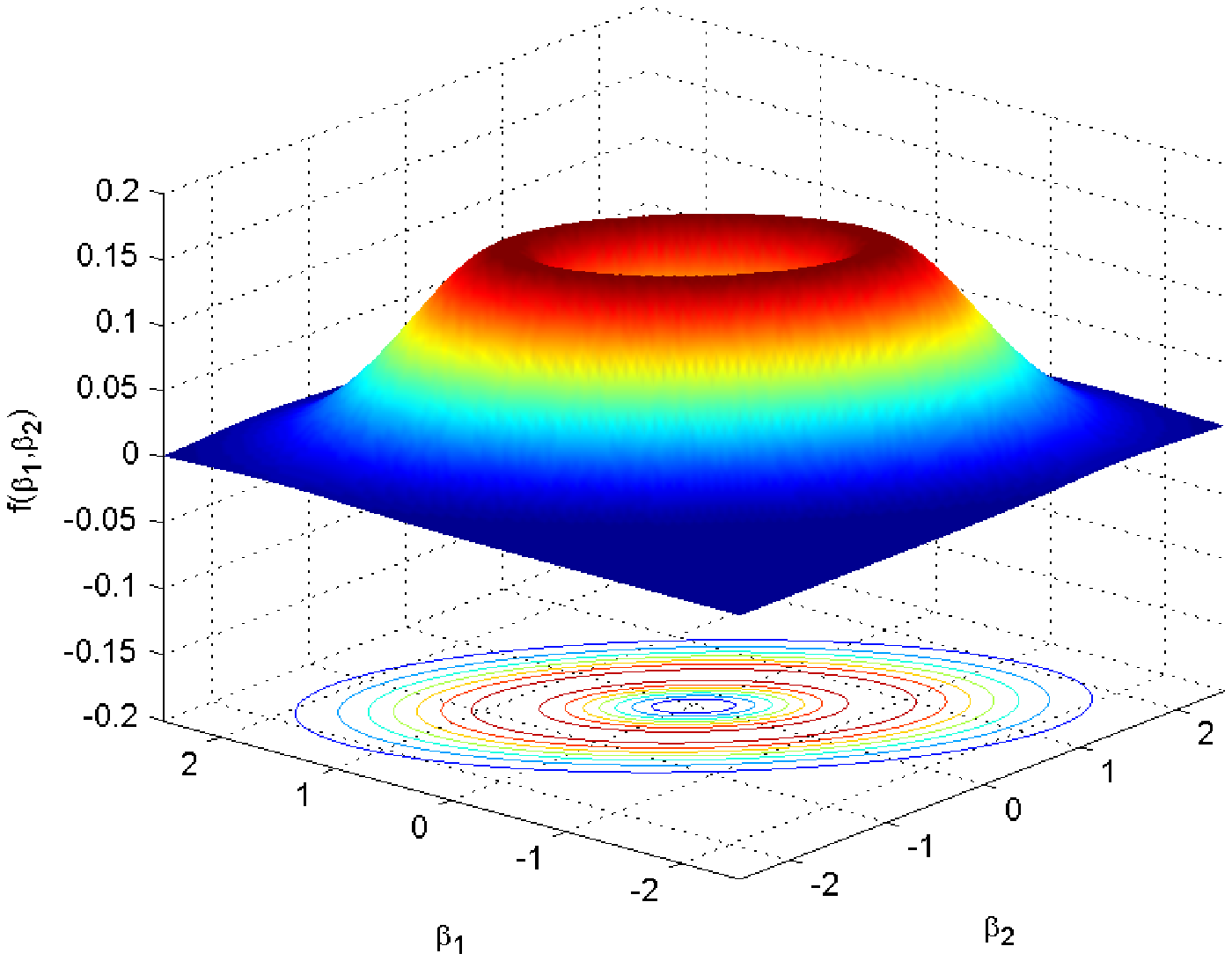}}
  \caption{ Bivariate quadratic normal moment prior with $\tau\sigma^2=0.1$ and $\bs{\Sigma} = (\bs{S}^\prime \bs{S})^{-1}$, where $\bs{S}=[\bs{S}_1,\bs{S}_2]^\prime$, $\bs{S}_1=(S_{1,1},S_{1,2},\dots, S_{1000,1})^\prime$, $\bs{S}_2=(S_{1,2},S_{2,2},\dots, S_{1000,2})^\prime$ and $ S_{i1} \in\{0, 1, 2\}$ and $ S_{i2} \in\{0, 1, 2\}$.  We introduce correlation between $S_{i1}$ and $S_{i2}$ through the latent variables $(Z_{i1}, Z_{i2})$, where $Z_{i1} \iid \textsf{N}_1(0,1)$, $Z_{i2} \iid \textsf{N}_1(0,1)$ and $\textsf{Cov}(Z_{i1}, Z_{i2}) = \rho$.  let $S_{i1} = 0$ if $Z_{i1} \le C_0$;  $S_{i1} = 2$ if $Z_{i1} > C_1$; and $S_{i1} = 0$ otherwise with $C_0=\Phi^{-1}((1-p_a)^2)$ and $C_1=\Phi^{-1}\left(1-p_a^2\right)$ where $\Phi^{-1}(\cdot)$ denotes normal inverse cumulative distribution function (CDF). We consider four scenarios when $p_a=0.8$:  (a) $\rho=0$; (b) $\rho=0.25$; (c) $\rho=0.5$; (d) $\rho=0.75$ with contours drawn beneath the PDF's surface. \label{fig:bi_QNMGP}}
\end{figure}

A prior distribution for $\bs{\beta}$ is needed to calculate the marginal likelihood of the data under $\textsf{H}_1$, for which we use the QNM prior with the density 
\begin{align*}
f_{\textsf{QNM}}(\bs{\beta}; \tau, \sigma^2,  \bs{\Sigma}) &= 
\frac{\bs{\beta}^\prime  \bs{\Sigma}^{-1}\bs{\beta}}{I\tau\sigma^2p}
f_{\textsf{N}_{p}}(\bs{\beta}; \bs{0}, I\tau\sigma^2\bs{\Sigma}),
\end{align*}
where $f_{\textsf{N}_{p}}(\cdot; \bs{m}, \bs{V})$ is the $p$-dimensional multivariate normal distribution with the mean vector $\bs{m}$ and covariance matrix $\bs{V}$, and $\tau$ is the dispersion parameter. As shown in the left panel of Figure \ref{fig:uni_QNMGP}, given $\sigma^2$ and $\bs{\Sigma}$, the bigger the $\tau$, the larger the mode and dispersion of the prior. The QNM prior increases the evidence in favor of both the true null and true alternative hypothesis, compared to other prior distributions (e.g. intrinsic and Cauchy priors).\cite{johnson2010use} Moreover, we specify $\sigma^2\bs{\Sigma}$ as the covariance matrix of the (iterative weighted) least square estimation of $\bs{\beta}$ in the GLM. This choice not only leads to convenient computation but also easily incorporates the prior knowledge about the effect of local ancestry on the trait. For example, when $\bs{S}_i$ is orthogonal to $\bs{E}_i$, $\bs{\Sigma} = (\bs{S}^\prime \bs{S})^{-1}$ with $\bs{S}=[\bs{S}_1, \bs{S}_2,\dots, \bs{S}_I]^\prime$ in the linear model for the continuous trait. As illustrated by the right panel of Figure \ref{fig:uni_QNMGP}, the QNM prior with $\bs{\Sigma} = (\bs{S}^\prime \bs{S})^{-1}$ suggests that for each locus, the higher the proportion of alleles from the high risk population ($p_a$), on average the larger the risk effect of local ancestry. Such relationships are frequently observed in admixture mapping. More importantly, when we investigate multiple loci simultaneously, it is crucial to take the correlation (linkage disequilibrium, LD) between the local ancestries into consideration. Figure \ref{fig:bi_QNMGP} plots several volcano-shaped bivariate QNM densities for various correlations between two local ancestries. It is clear that for two loci with admxiture linkage equilibrium (as shown in panel (a)), such as two loci on different chromosomes, their risk effects would be independent;  and that for two loci with high admixture LD (as shown in panel (d)), usually located in the same gene, they would have similar risk effects. 

We use the Bayes factor to compare the likelihoods of observed traits under $H_1$ versus under $H_0$. Intuitively, the Bayes factor is the ratio between the evidences which combine the likelihood of the observed traits with the prior probability of association under the $H_1$ and $H_0$ respectively. The larger the Bayes factor, the stronger the evidence would be in support of $H_1$.    
With QNM prior for $\bs{\beta}$ under $\textsf{H}_1$, the Bayes factor can be obtained in the simple closed form,
\begin{equation}
\label{eq:BF}
BF(\bs{\mathrm{y}}) =
\frac{p+\bs{T}}{p(1+I\hat{\tau})^{p/2+1}}
\exp\left(\frac{\bs{T}}{2}\right),
\end{equation}
where $\bs{T}=\frac{I\tau}{\hat{\sigma}^2(1+I\hat{\tau})}{\bs{\hat{\beta}}}^\prime{\bs{{\widehat{\Sigma}^{-1}_\beta}}}{\bs{\hat{\beta}}}$, 
$\bs{\hat{\beta}}$ is the maximum likelihood estimate of $\bs{\beta}$, adjusted by other risk covariates when necessary, $\bs{{\widehat{\Sigma}^{-1}_\beta}}$ is the corresponding covariance matrix estimates and $\hat{\tau}$ and $\hat{\sigma}^2$ are the empirical Bayes estimates. Bayes factor \eqref{eq:BF} will be used to identify the loci associated with the traits, detailed as follows.
   
\subsection*{Generalized admixture mapping procedure}
We propose a two-stage approach for GLEAM. In the first stage, we examine the marginal association between a single AIM locus and the trait, using the Bayes factors \eqref{eq:BF}, one locus a time for $J$ AIM loci. The loci at which $log_{10} BF(\bs{\mathrm{y}}) > \delta$ are considered susceptibility loci. While the `one locus a time' approach explores the marginal association and is widely used, marginal association only reflects part of the relationship between the AIM loci and the trait.
Several loci in different regions may show associations with the trait. Thus, it is desirable to quantify the evidence for joint association of multiple loci with the trait. For this reason, in the second stage, we list all possible combinations of susceptibility loci selected in the first stage. For each set of susceptibility loci, we can again calculate the Bayes factors for the joint association at those loci simultaneously. The most significant ones are reported. The local ancestries at the AIM loci are unobserved and imputed from the HMM. The imputation uncertainty could be properly accounted for by calculating weighted average of the Bayes factors for each imputed local ancestry dataset, which is similar to the strategy used by Guan and Stephens \cite{guan2008practical} in imputation-based association mapping for testing untyped variants.  

\subsection*{Simulation Studies}
We carried out simulation studies to assess the performance of GLEAM in terms of type I error rate and power under various scenarios and compared it with the method based on Bayesian likelihood ratio (BLR) by Patterson et al.  \cite{patterson2004methods} which is implemented by the software ANCESTRYMAP (\urlhttp genepath\urldot me\urldot harvard\urldot edu/\urltilda reich/Software\urldot htm). GLEAM and ANCESTRYMAP use slightly different HMMs to impute the local ancestries and ANCESTRYMAP records the proportion of local ancestries only. Because of these differences, we assumed the true local ancestries were given and focused on evaluating the ability of localizing susceptibility loci, instead of estimating local ancestries. Our simulations were based on empirical data of local ancestries for 1001 African Americans from the HPHB Study \cite{miranda2009environmental},  with 1296 AIM loci measured across the genome.

We started by investigating the type I error rates for the local ancestries which were scattered around different regions of the genome and in linkage equilibrium. Under this scenario, the falsely localized AIM locus would be in the region remote from the true disease causing locus, which leads to a false positive finding. We first randomly sampled 1000 AIM loci with replacement from 1296 AIM loci for 1000 subjects. At each AIM locus, we simulated the local ancestries measured by the number of alleles from the African ancestral population from their maximum a posteriori (MAP) frequency estimates under the assumption of Hardy-Weinberg equilibrium. Ten sets of trait data were then generated such that we were able to assess the type I error rates under the genome-wide threshold level (e.g $\alpha = 10^{-4}$), by using the following null model for continuous traits:
\begin{equation*}
\mathrm{y}_i=\alpha E_i + \varepsilon_i,
\end{equation*}
and for binary traits,
\begin{equation*}
\textsf{logit}\{\textsf{Prob}(\mathrm{y}_i=1)\}= \alpha E_i ,
\end{equation*}   
where the continuous risk covariate $E_i$ and the measurement error $\varepsilon_i$ followed standard normal distributions. We considered two situations whereby $\alpha=0$ in the absence of a covariate effect and $\alpha=1$ in the presence of a covariate effect.

We next examined power under the single locus alternative models. We simulated 100 sets of traits. Each set included 1000 subjects and one disease associated local ancestry whose location was randomly sampled from 259 AIM loci, where the proportion of African ancestral population (PAAP) ranged from $0.8321$ to $0.8817$ and was on the top $20\%$ percentile among 1296 AIM loci. Given the local ancestry $S_i$, continuous covariates $E_i$ and measurement error $\varepsilon_i$ generated same as that for the null model, continuous traits were simulated from 
\begin{equation*}
\mathrm{y}_i=\alpha E_i + \beta S_i + \varepsilon_i,
\end{equation*}
and binary traits from,
\begin{equation*}
\textsf{logit}\{\textsf{Prob}(\mathrm{y}_i=1)\}= \alpha E_i+ \beta S_i.
\end{equation*}
Under both models, the $\beta$ was specified as  $\beta=c\times\textsf{PAAP}$ which reflected the \emph{a priori} observation that the locus with the larger proportion of the high risk ancestral (here African American) population usually demonstrated stronger association with the traits. For continuous traits, we chose the values of effect size multiplier $c$ as 0.2, 0.25, 0.3, 0.35 and 0.4 respectively, with the largest possible effect size equal to 0.3527. Similarly, we picked the values of $c$'s as 0.4, 0.5, 0.6, 0.7 and 0.8  for binary traits with the largest possible odds ratio (OR) equal to $1.8537$. 

We further considered a multilocus alternative model where two local ancestries were associated with the traits and there existed admixture linkage disequilibrium. To do so, we generated an artificial chromosome composed of two pieces from chromosome 1 and chromosome 4 with the length 139.50Mb and 114.88Mb respectively for 1000 subjects, based on empirical data on local ancestries from HPHB study. In the middle of each chromosome piece with 51 loci, there is one locus whose proportion of African ancestry population was among the highest in all 1296 AIM loci. In the simulations, those two loci are assumed to be associated with traits. We generated 100 sets of continuous and binary traits respectively, each of which was simulated similarly to the single locus alternative model except with two local ancestries involved and both effect size multiplier $c$'s set at 0.7 for continuous traits and 0.35 for binary traits.            

The simulated datasets were analyzed by the GLEAM and the  BLR method. Since the BLR method was primarily developed for binary traits, the BLR method required transformation of continuous traits into binary ones, such as defining the subjects with top $20\%$ traits as the cases and the one with bottom $20\%$ traits as controls. 

\section*{Results}
\subsection*{Simulation Studies}
\begin{figure}
  \centering
\subfigure[Generalized admixture mapping ]{\includegraphics[width=0.48\textwidth]{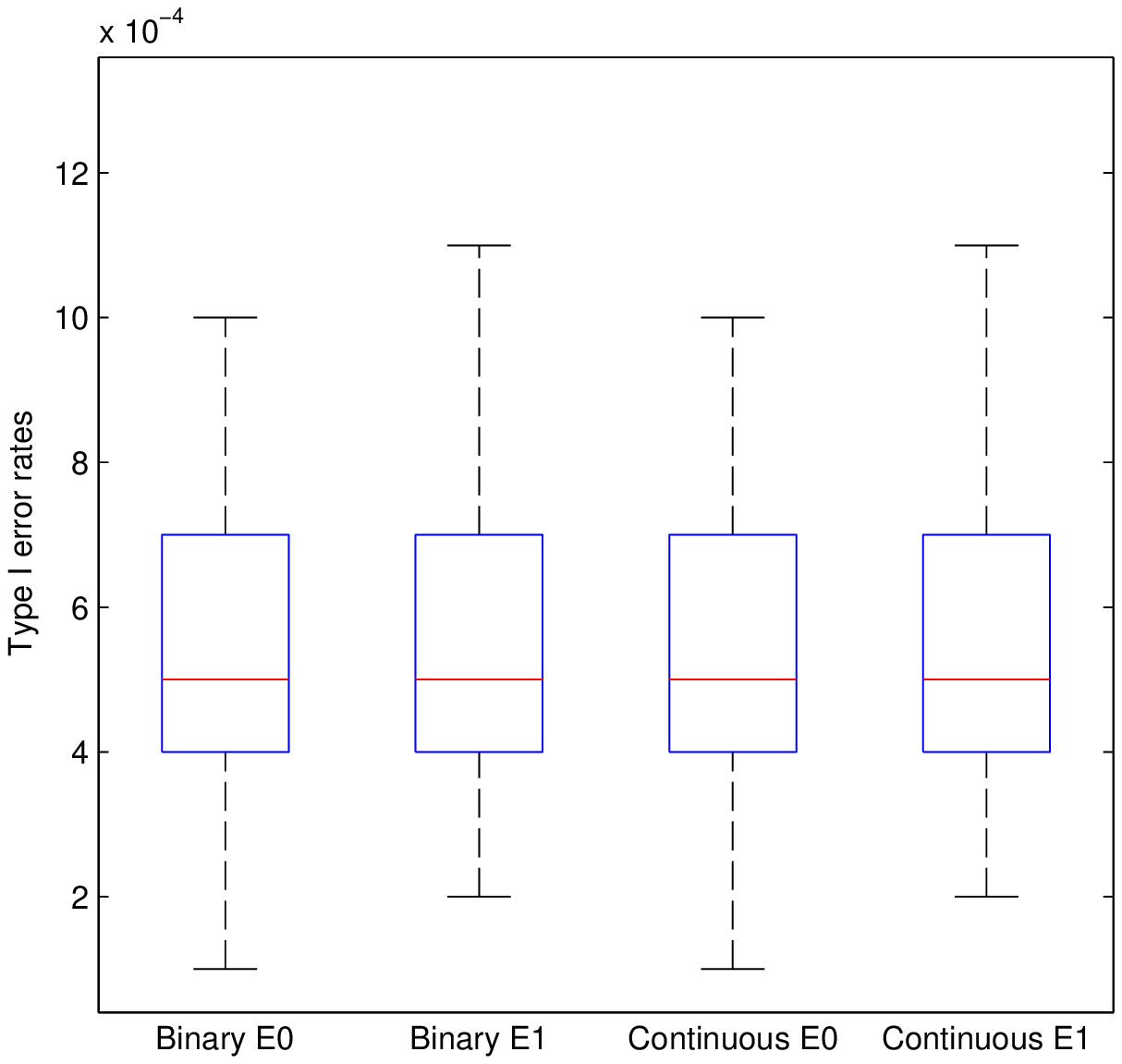}}
\subfigure[Method based on BLR]{\includegraphics[width=0.48\textwidth]{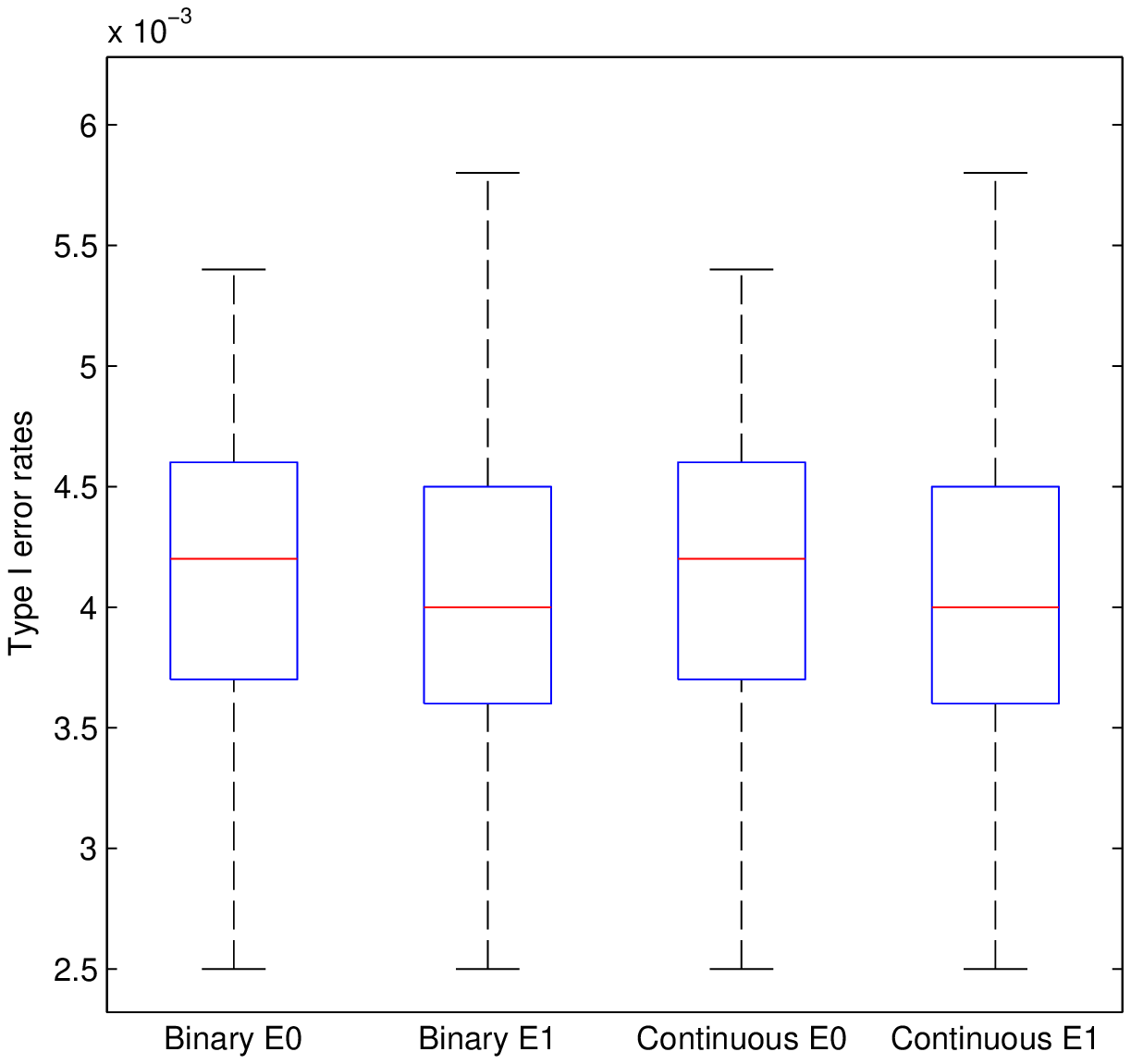}}
  \caption{The type I error rates under the null model (Note the different scaling of the Y-axis for panels a and b). The type I error rates are presented for both the binary and continuous traits respectively, with or without covariate effect. For each simulated dataset, we calculate one type I error rate under the genome-wide threshold level 2 for both methods. The results for 100 replications are summarized by the boxplots, where the center bar is median, bottom and top of the box are the 25th and 75th percentile and the whiskers stretch out till the extreme values.     
\label{fig:nullmodel}}
\end{figure}

Figure \ref{fig:nullmodel} presents the empirical type I error rates for both the binary and continuous traits, with or without covariate effects. For the GLEAM and the BLR methods, we chose a threshold of 2 for $log_{10} BF(\bs{\mathrm{y}})$ to control the genome-wide type I error rates. Under the null model that all the local ancestries are in linkage equilibrium, the type I error rate is controlled at a low level with the median around $5\times10^{-4}$ for GLEAM and $4.2\times10^{-3}$ for the BLR method illustrated in Figure \ref{fig:nullmodel}. In both cases, those type I error rates seem overly conservative. However, in the application to real data, slight admixture linkage disequilibrium between the AIM loci will significantly inflate the type I error rate close to the nominal levels (i.e. $\alpha = 0.05$ or $0.005$), which is discussed in the later paragraphs. Comparing two panels in Figure \ref{fig:nullmodel} reveals that the type I error rates of GLEAM are consistently smaller than those of the method based on BLR and are little affected by the presence of covariate effects when properly adjusted. The covariates are not considered by the BLR method and have a mixed effect on type I error rates, where the median is slightly reduced with the maximal type I error rates increased.          

\begin{figure}
  \centering
\subfigure[Binary traits without covariate effect ]{\includegraphics[width=0.48\textwidth]{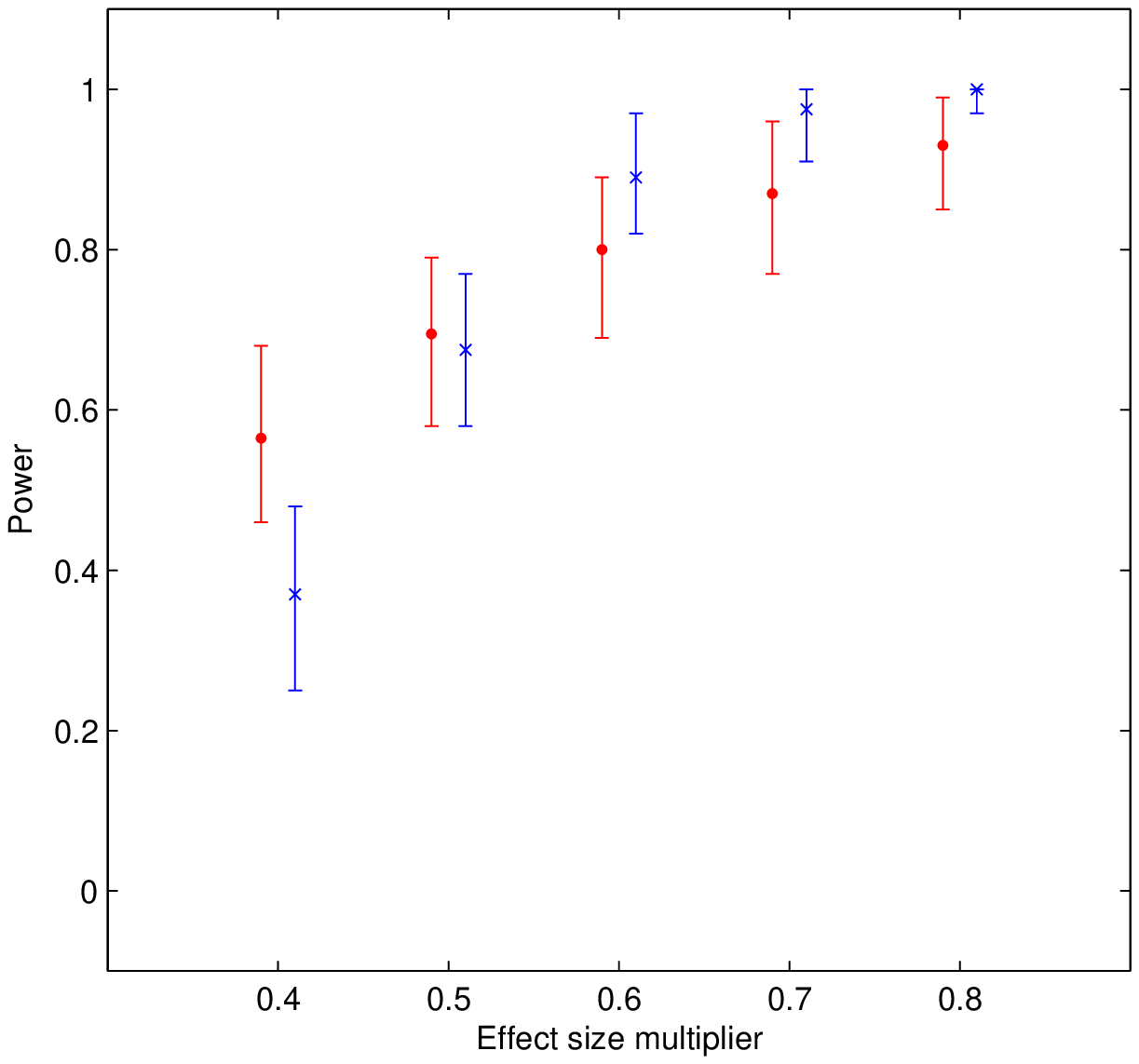}}
\subfigure[Binary traits with covariate effect ]{\includegraphics[width=0.48\textwidth]{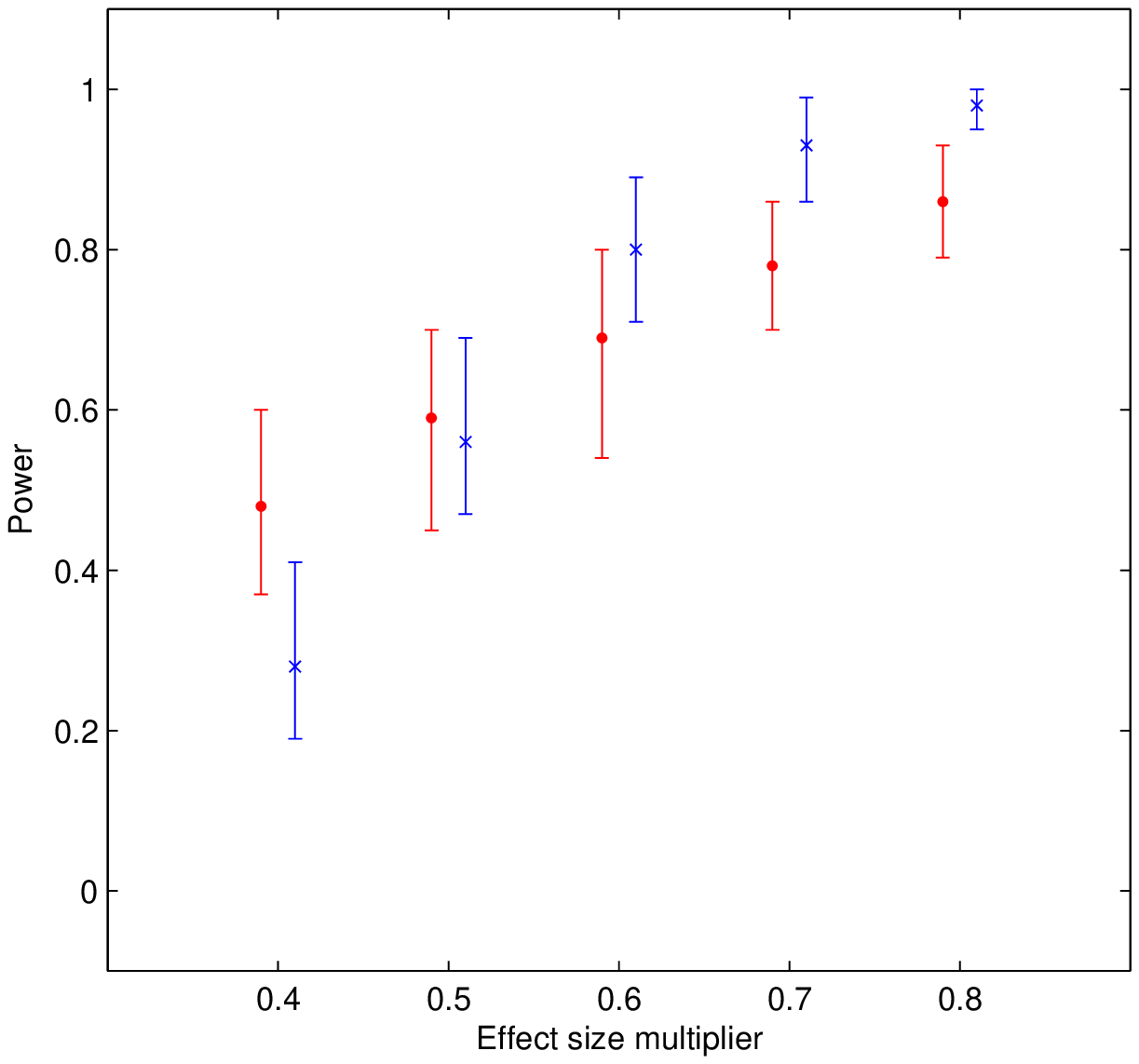}}\\
\subfigure[Continuous traits without covariate effect ]{\includegraphics[width=0.48\textwidth]{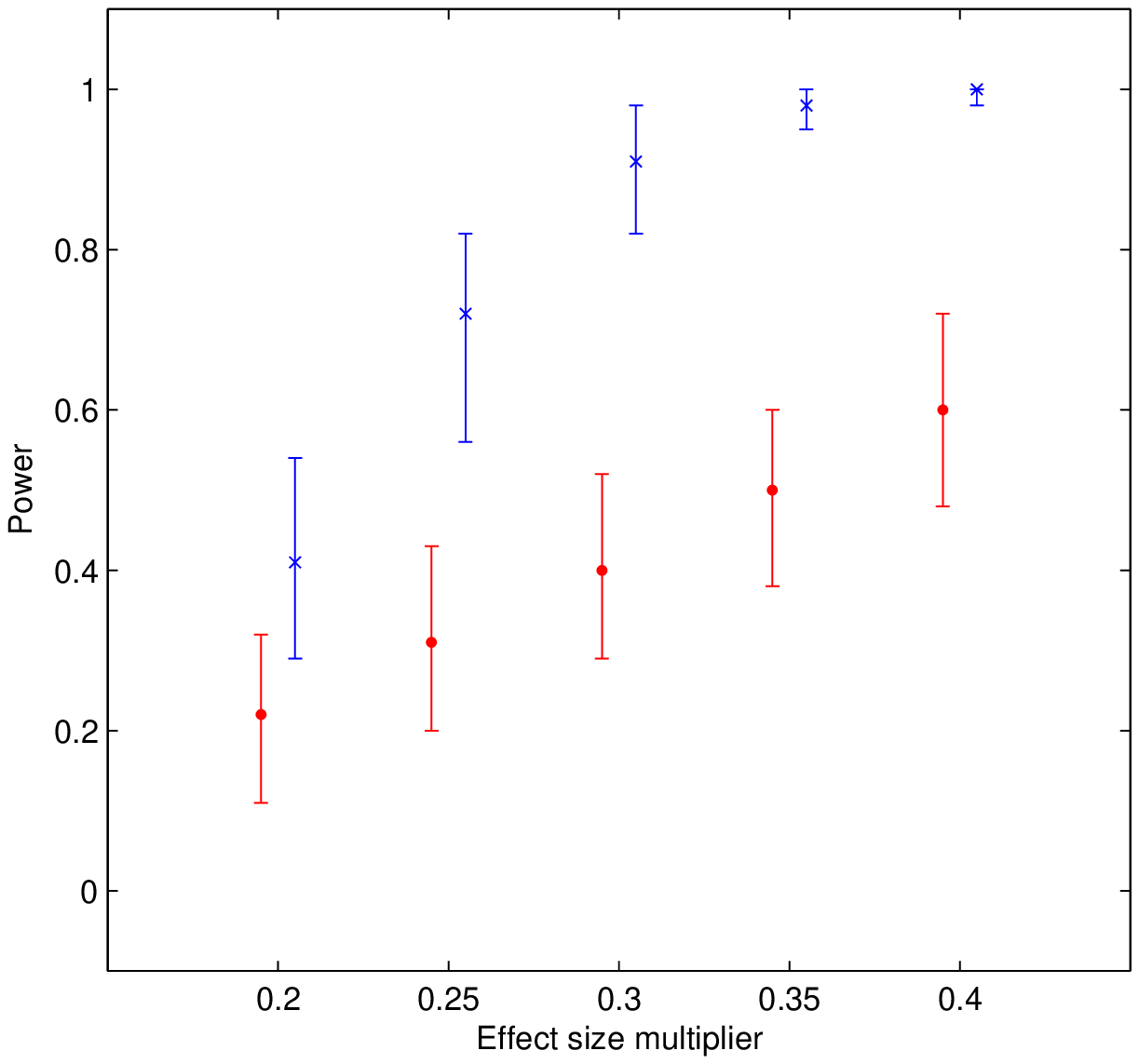}}
\subfigure[Continuous traits with covariate effect ]{\includegraphics[width=0.48\textwidth]{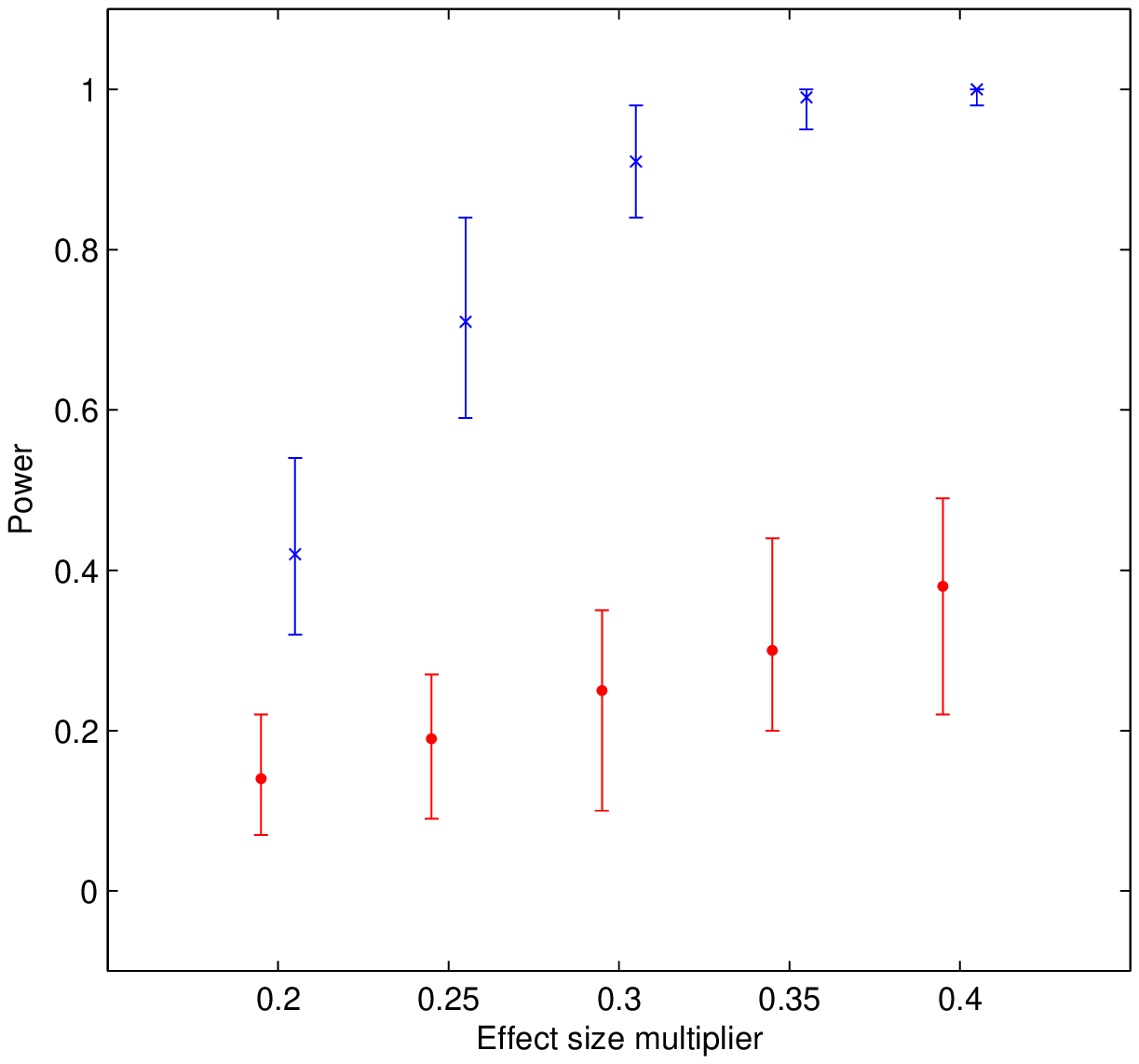}}
  \caption{Powers for single locus alternative models. Power is calculated for each dataset with 100 replications total for the binary or continuous traits simulated under the single locus alternative model with or without covariate effect. The {\color{blue}$\times$} indicates the median of powers by the GLEAM and {\color{red} \textbullet }\;denotes the median of powers by the method based on Bayesian likelihood ratio. The whiskers on each bar represent the minimal and maximal powers respectively. The effect sizes of local ancestries are equal to the multiplication of effect size multiplier $c$ and the proportion of African ancestry population. \label{fig:singleAlt}}
\end{figure}

Power of the methods was also evaluated for binary and continuous traits under the single locus alternative model, with or without covariate effects. We considered various effect sizes of local ancestries with the results shown in Figure \ref{fig:singleAlt}. For the binary trait, when the effect size is small, the BLR method performs better with larger power. With the increment of the effect sizes, GLEAM gradually outperforms the BLR method. For both methods, covariates have moderate effects on power, which is more obvious for the smaller effect sizes. For the continuous trait, the GLEAM performs significantly better at each effect size. These results were expected since the BLR method discards part of the dataset in order to transform the continuous trait into the binary one (case versus control), which inevitably loses power. For all situations considered, the power of the GLEAM approach increases with the increment of the local ancestry effect size, most rapidly when the effect sizes are smaller and then levels off with larger effect sizes. In comparison, the power of the BLR method increases roughly linearly.
        
To understand the impact of admixture linkage disequilibrium on type I error rates and to evaluate the ability of localizing multiple loci simultaneously, we generated a set of artificial chromosomes as described before, where two loci were associated with the traits, named as Locus 1 and Locus 2. Besides Locus 1 and Locus 2, we divided the remaining loci into three regions: region 1 (REG1) with 42 loci and region 2 (REG2) with 35 loci, where the admixture linkage disequilibrium measured by the correlation coefficient between a given locus at these regions and Locus 1 or Locus 2 was larger than 0.12 respectively; and region 3 (REG3), the unassociated loci which did not belong to region 1 and region 2. Strictly speaking, the identified loci except Locus 1 and Locus 2 were all false positives. However, in contrast to the loci found in region 3 which were completely false findings, the loci identified in Region 1 and Region 2 were partially correct and could be regarded as low resolution findings instead, since the true associated locus did exist in the nearby region. Therefore, we evaluated the false positives in three regions separately. An ideal method under the pre-specified genome-wide threshold would lead to few completely false positives in region 3 and to a small number of partially false positives in regions 1 and 2, while being able to identify the true associated loci with high frequency. 

Table \ref{tbl:multiAlt} summarizes the frequencies of identified loci for each locus or locus combination at different regions by GLEAM and BLR method. For the GLEAM method, we applied the two-step approach outlined in the ``Generalized admixture mapping procedure'' subsection. The results by applying the first step only (GLEAM1) and by applying the two-step approach (GLEAM2) were both presented. For binary traits, both the BLR method and GLEAM1 could localize both Locus 1 and Locus 2 with high power.
The type I error rates in region 1 were around the nominal level (0.025 and 0.003 respectively). The type I error rates in region 1 and region 2 were higher than the ones in region 3, which would decrease the resolution of the finding. Compared to GLEAM1, further applying the second step of generalized admixture mapping procedure (GLEAM2) could significantly improve the resolution by reducing the type I errors in region 1  (from 0.013 to 0.002) and region 2 (from 0.014 to 0.003). For continuous traits, GLEAM2 also performed best with much higher power and lower type I rate than the BLR method.  
  
\begin{table}
\caption{The frequency of identified loci for each locus or locus combination at different regions of the artificial chromosome.\label{tbl:multiAlt} }
{\footnotesize
\begin{center}
\begin{threeparttable}
\begin{tabular}{lccccccc}
\toprule
Trait&Method&REG1&REG2&REG3&Locus1&Locus2&Locus1/2\tnote{a}\\
\toprule\\
 & BLR  & 0.103 & 0.047 & 0.025 & 0 & 0 & 1.000 \\
\cmidrule{2-8}\\
Binary&GLEAM1\tnote{b}& 0.013 & 0.014 & 0.003 & 0.020 & 0.020 & 0.960\\
\cmidrule{2-8}\\
&GLEAM2\tnote{c}& 0.002 & 0.003 & 0.001 & 0.030 & 0.030 & 0.940 \\
\midrule\\
 & BLR  &  0.035 & 0.018  &  0.011 & 0.030 & 0.400 & 0.560 \\
\cmidrule{2-8}\\
Continuous&GLEAM1& 0.021 & 0.017 & 0.004 & 0.030 & 0 & 0.970\\
\cmidrule{2-8}\\
&GLEAM2& 0.004 & 0.003 & 0.002 & 0.040 & 0 &  0.960\\
\bottomrule
\end{tabular}
 \begin{tablenotes}[online]
   \item \hspace{3pt} a: The combination of Locus 1 and Locus 2
   \item \hspace{3pt} b: Applying the first step of generalized admixture mapping procedure only;
   \item \hspace{3pt} c: Applying both steps of generalized admixture mapping procedure;
   \item \vspace{-5pt} \hspace{2pt} \line(1,0){333}
   \item \vspace{-8.6pt} \hspace{2pt} \line(1,0){333}
   \item \vspace{-8.7pt} \hspace{2pt} \line(1,0){333}    
\end{tablenotes}
\end{threeparttable}
\end{center}
}
\end{table}
\subsection*{Application}
This methodological work was motivated by real data from the Healthy Pregnancy, Healthy Baby (HPHB) study, which is a prospective cohort study of pregnant women aimed at identifying genetic, social and environmental contributors to disparities in adverse birth outcomes in the US south.\cite{miranda2009environmental} Consistent with previous studies, African American women in HPHB have higher risk for maternal hypertension than Caucasian women during the pregnancy, which contributes to the poor birth outcomes\cite{allen2004effect}. Even within the African American subpopulation, some African American women have much higher blood pressures, and we hypothesize that one possible contributor may be the percentage of African ancestry. To explore this hypothesis, we applied GLEAM to investigate the association between the averaged maternal mean arterial pressure (MAP), defined as $(1/3\times \text{systolic blood pressure})+(2/3 \times \text{diastolic blood pressure})$,  during 24 to 28 weeks of pregnancy and local ancestries among these pregnant African American women. 
Clinical and genetic data were available for 1004 nonHispanic Black (NHB) women. 1509 SNP AIMs were genotyped using the Illumina African American admixture panel. After quality control measures described previously \cite{Ashley2011}, the dataset consisted of 1001 NHB women with 1296 AIMs.

\begin{figure}[h!]
  \centering
    \includegraphics[width=0.9\textwidth,angle=0]{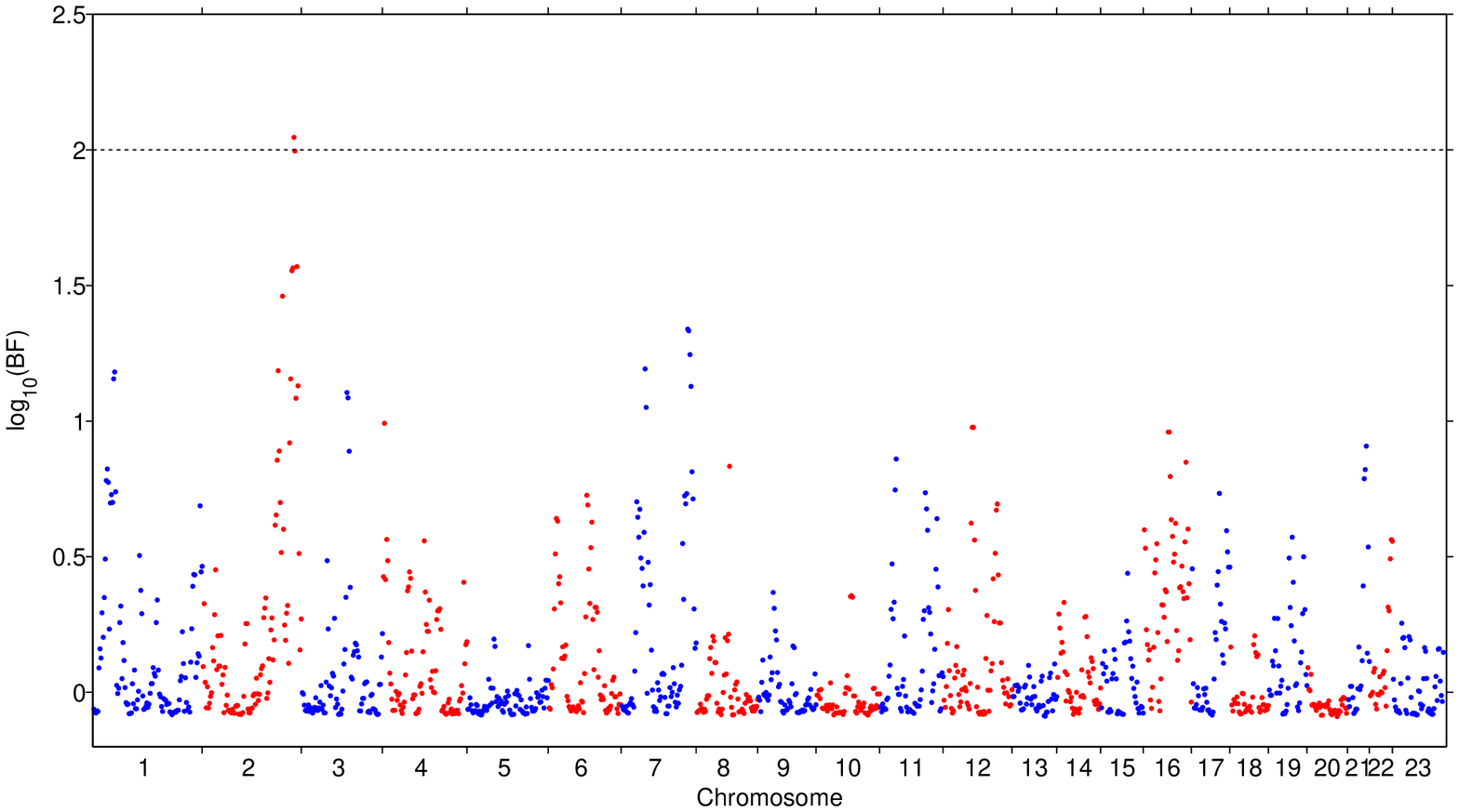}
  \caption{Manhattan plot of $log_{10}\text{(Bayes factor)}$ for the association between the averaged maternal mean arterial pressure (MAP) during 24 to 28 weeks of pregnancy and genome-wide local ancestries among 1001 African Americans. \label{fig:gest_aged}}
\end{figure}

The proposed GLEAM approach was applied to this dataset to identify the local ancestry associated with the averaged maternal MAP, a continuous trait, while adjusting for mother's age. The local ancestries were multiply imputed based on the HMM.  We first examined the marginal association between the trait and local ancestries, one locus a time. The results were summarized in Figure \ref{fig:gest_aged},  where one local ancestry on the chromosome 2 was identified with its $log_{10}\text{(Bayes factor)}=2.05$ exceeding the threshold 2. With only one local ancestry localized, the second step of the generalized admixture mapping procedure was unnecessary. The same data were analyzed by the BLR method, which treated the subjects with averaged maternal MAP more than 93.67 (top $20\%$ quantile) as cases and the ones with averaged maternal MAP less than 79.33 (bottom $20\%$ quantile) as control. No local ancestry was identified as being associated with the averaged maternal MAP with this approach, presumably due to its relatively low power compared with the GLEAM approach.
 
\section*{Discussion}
By utilizing admixture linkage disequilibrium, admixture mapping is an indispensable tool to localize the alleles which are associated with the qualitative or quantitative traits and diseases that vary in prevalence across the ancestral populations. The available methods are most suitable for dichotomous traits in a case-control study and do not allow for adjustment for other risk covariates. In this article, we propose a flexible and powerful generalized admixture mapping approach, which is based on the generalized linear model and is able to incorporate admixture prior information by using the quadratic normal moment prior and to adjust for covariates. The proposed method is applicable to both qualitative and quantitative traits with satisfactory power while controlling the type I error rates at a low level, and is able to be easily implemented as we demonstrated with our HPHB example.  

In addition to the flexibility to handle different types of traits, other attractive generalizations include consideration of multiple loci simultaneously. As illustrated in Figure \ref{fig:ALD}, admixture linkage disequilibrium extends much further than  haplotype linkage disequilibrium. Consequently, if we only examine one locus a time, the local ancestries which are highly correlated to the true disease associated local ancestry tend to be identified as significant ones as well. As demonstrated by the simulations, those false positives can be significantly reduced by considering multiple susceptible loci simultaneously, which reduce the type I error rates and improve the mapping resolution. In addition, GLEAM specifies a hidden Markov model treating the recombination rates varying across the genome, which allows us to infer the recombination ``hotspots'' in admixture population. Moreover, within the generalized linear model framework, it is straightforward to extend the current method to populations with more than to two ancestral populations, such as Hispanic populations, by adding extra ancestry population covariates. It is also easy to consider the interaction between the local ancestries and covariates with the properly specification of the priors on interaction coefficients.         
  
\section*{Acknowledgments}
This work was supported by Award Number R01ES017436 from the National Institute of Environmental Health Sciences, and by funding from the National Institutes of Health (5P2O-RR020782-O3) and the U.S. Environmental Protection Agency (RD-83329301-0). The content is solely the responsibility of the authors and does not necessarily represent the official views of the National Institute of Environmental Health Sciences, the National Institutes of Health or the U.S. Environmental Protection Agency.

\newpage
\section*{Appendices}
\subsection*{MCMC algorithm for HMM}
We propose an MCMC algorithm for posterior computation of HMM as follows. 
\begin{enumerate}[(a)]
\item Impute the missing AIM $X^m_{ij}$. Given the $\bs{P}_j$ and $S_{ij}$, $X^m_{ij} \in \{0,1,2\}$ can be easily sampled with probability mass $p_j(S_{ij},X^m_{ij})$.  

\item Update the latent states $\bs{S}_i$ for $i=1,2,\cdots, I$. Given the $\bs{Q}_{i0}$,  $\bs{Q}^{(r)}_i$ and $\bs{R}_i=\{R_{ij}\}_{1 \times J}$, we will use the forward filtering backward sampling (FFBS) algorithm \cite{scott2002bayesian} to sample the $\bs{S}_i$ in one block. The FFBS algorithm mixes more rapidly comparing to the direct Gibbs sampler which samples one $S_{ij}$ a time conditional on the remains of $\bs{S}_{i}$. Let $\bs{X}^j_{i1}=\left[X_{i1},X_{i2},\cdots, X_{ij}\right]'$ and  $\bs{R}_i=\left[R_{i1},R_{i2},\cdots, R_{iJ}\right]'$. We begin the FFBS algorithm by calculating $\bs{Q}^F_{ij}=\{q^F_{ij}(m,n)\}_{3 \times 3}$ with $q^F_{ij}(m,n) = \Prob(S_{i(j-1)}=m, S_{ij}=n \mid \bs{X}^j_{i1}, \bs{R}_i)$ recursively for $j = 1, 2, \cdots, J$ as
\begin{align*}
q^F_{ij}(m,n) 
&=\Prob(S_{i(j-1)}=m, S_{ij}=n \mid \bs{X}^j_{i1}, \bs{R}_i) \nonumber\\ 
&=\frac{\Prob(S_{i(j-1)}=m, S_{ij}=n,  X_{ij}\mid \bs{X}^{j-1}_{i1}, \bs{R}_i)}{\Prob(X_{ij}\mid \bs{X}^{j-1}_{i1}, \bs{R}_i)} \nonumber\\
&=\frac{q^F_{i(j-1)}(m)q^{(r)}_{i}(m,n)p_{j}(n,X_{ij})}{\Prob(X_{ij}\mid \bs{X}^{j-1}_{i1}, \bs{R}_i)},\label{eq:likelihood}
\end{align*} 
where $q^F_{i0}(m)=\bs{Q}_{i0}$, $\Prob(X_{ij}\mid \bs{X}^{j-1}_{i1}, \bs{R}_i) = \sum^2_{m=0}\sum^2_{n=0} \Prob(S_{i(j-1)}=m, S_{ij}=n,  X_{ij}\mid \bs{X}^{j-1}_{i1}, \bs{R}_i)$,
and $q^F_{ij}(n)=\sum^2_{m=0}q^F_{ij}(m,n)$. 
   
We can then sample the $\bs{S_i}$ backward from $S_{iJ}$ to $S_{i1}$ with 
\begin{equation*}
\Prob(\bs{S}_i \mid \bs{X}_i, \bs{R}_i) = \Prob(S_{iJ} \mid \bs{X}_i, \bs{R}_i) \prod^{J-1}_{j=1}\Prob(S_{i(J-j)} \mid \bs{S}^J_{i(J-j+1)}, \bs{X}_i, \bs{R}_i),
\end{equation*}     
where
\begin{align*}
\Prob(S_{iJ} \mid \bs{X}_i, \bs{R}_i) &= q^F_{iJ}(S_{iJ}), \\
\Prob(S_{i(J-j)} \mid \bs{S}^J_{i(J-j+1)}, \bs{X}_i, \bs{R}_i) 
&=\Prob(S_{i(J-j)} \mid S_{i(J-j+1)}, \bs{X}^{J-j+1}_{i1}, \bs{R}_i) \\
&=\frac{q^F_{i(J-j+1)}(S_{i(J-j)}, S_{i(J-j+1)})}{q^F_{i(J-j+1)}(S_{i(J-j+1)})}.
\end{align*}
The initial state $S_{i0}$ will be sampled with $\Prob( S_{i0} \mid \bs{S}_i, \bs{X}_i, \bs{R}_i) = \frac{q^F_{i1}(S_{i0}, S_{i1})}{q^F_{i1}(S_{i1})}$.

\item Update the recombination count $\bs{R}_i =\{R_{ij}\}_{1\times J}$ for $i=1,2,\cdots,I$. $R_{ij}$ is sampled with  full conditional probability mass function
\begin{equation*} 
\Prob(R_{ij} \mid S_{i(j-1)}=m, S_{ij}=n, \bs{Q}^{(0)}_i,\bs{Q}^{(1)}_i,\bs{Q}^{(2)}_i, \gamma_j) = \frac{q^{(R_{ij})}_i(m,n){2 \choose R_{ij}} \gamma^{R_{ij}}_j(1-\gamma_j)^{2-R_{ij}}}{\sum^2_{r=0}q^{(r)}_i(m,n){2 \choose r} \gamma^r_j(1-\gamma_j)^{2-r}}
\end{equation*}

\item Update recombination probability $\gamma_j$ from $\textsf{Beta}\left( \tau^\gamma \gamma_{0j} + \sum^I_{i=1} R_{ij}, \tau^\gamma (1-\gamma_{0j})+ 2I - \sum^I_{i=1} R_{ij}  \right)$ for $j=1,2,\cdots, J$.

\item Update the proportion ancestry from population A $\rho_i$ from \\ $\textsf{Bin}
\left(
\tau^\rho \rho_{0i}+n^{(1)}_{01}+n^{(1)}_{12}+n^{(1)}_{22}+n^{(2)}_{\cdot 1}+2n^{(2)}_{\cdot 2}, 
\tau^\rho (1-\rho_{0i})+n^{(1)}_{00}+n^{(1)}_{10}+n^{(1)}_{21}+n^{(2)}_{\cdot 1}+2n^{(2)}_{\cdot 0}
\right)$, where $n^{(1)}_{kl}=\sum^J_{j=1}I(S_{i(j-1)}=k \text{ and } S_{ij}=l \text{ and } R_{ij}=1)$ and $n^{(2)}_{\cdot l}=\sum^J_{j=1}I( S_{ij}=l \text{ and } R_{ij}=2)$.

\item Update $\bs{Q}^{(0)}_i$, $\bs{Q}^{(1)}_i$, $\bs{Q}^{(2)}_i$ and $\bs{Q}_{i0}$ based on last $\rho_i$ for $i=1,2,\cdots, I$.

\item Update $p^A_j$ and $p^B_j$ for $j=1,2,\cdots,J$.
Let $n_{kl}=\sum^I_{i=1}I(S_{ij}=k \text{ and } X_{ij}=l)$ and $n^{VA}_{11}$ denotes the case that the allele from population A is variant allele when $S_{ij}=1$ and $X_{ij}=1$. $n^{VA}_{11}$ is unobserved and can be imputed from $\textsf{Bin}\left(n_{11},\frac{p^A_j(1-p^B_j)}{p^A_j(1-p^B_j)+p^B_j(1-p^A_j)}\right)$.
$p^A_j$ is then sampled from $\textsf{Beta}\left(\tau^A p^A_{0j}+n_{21}+2n_{22}+n^{VA}_{11}, \tau^A(1-p^A_{0j})+n_{21}+2n_{20}+n_{11}-n^{VA}_{11} \right)$; $p^B_j$ is sampled from $\textsf{Beta}\left(\tau^B p^B_{0j}+n_{01}+2n_{02}+n_{11}-n^{VA}_{11}, \tau^B(1-p^B_{0j})+n_{01}+2n_{00}+n^{VA}_{11} \right)$

\item Update $\bs{P}_j$ based on last $p^A_j$ and $p^B_j$ for $j=1,2,\cdots,J$.

\item Update $\tau^A$ and $\tau^B$ using Random-Walk Metropolis-Hasting. For $\tau^A$, we propose the new $\tau^{A*} = \tau^A + \epsilon$ where $\epsilon \sim \textsf{N}_1(0, \sigma^2_{mh})$. The posterior distribution of $\tau^A$,  $f(\tau^A \mid \bs{p}^A ) \propto \prod^J_{j=1} f_{\text{Beta}}\left(P^A_j \mid \tau^A p^A_{0j}, \tau^A(1-p^A_{0j})\right)I(50<\tau^A<1000) $. Then, $\alpha(\tau^A, \tau^{A*})=\textsf{min}\left\{\frac{f(\tau^{A*} \mid \bs{p}^A )}{f(\tau^A \mid \bs{p}^A )}, 1 \right\}$. We draw $\mu^A \sim \textsf{U}[0,1]$. If $\mu^A < \alpha(\tau^A, \tau^{A*})$, then $\tau^A$ is replaced by $\tau^{A*}$; otherwise, $\tau^A$ is unchanged. Similar update is conducted for $\tau^B$.   
      
\end{enumerate}

\bibliographystyle{ajhg} 
\bibliography{gam}
\end{document}